\documentclass[a4paper,twocolumn,11pt, accepted=2023-04-17]{quantumarticle}
\pdfoutput=1
\usepackage[T1]{fontenc}
\usepackage[utf8]{inputenc}
\usepackage{amsmath}
\usepackage{hyperref}
\usepackage{braket}
\usepackage{graphicx}
\usepackage{cleveref}
\usepackage{nicefrac}
\usepackage{dsfont}
\usepackage{SIunits}
\usepackage[dvipsnames]{xcolor}
\usepackage{wasysym}
\usepackage[numbers,sort&compress]{natbib}
\usepackage{mathtools}
\usepackage{ulem}
\usepackage{bm}
\usepackage{enumitem}
\DeclarePairedDelimiter\ceil{\lceil}{\rceil}

\newtheorem{theorem}{Property}

\renewcommand{\emph}[1]{\textit{#1}}
\begin{document}

\title{Near-deterministic hybrid generation of arbitrary photonic graph states using a single quantum emitter and linear optics}
\author{Paul Hilaire}
\affiliation{Department of Physics, Virginia Tech, Blacksburg, Virginia 24061, USA}
\affiliation{Huygens-Kamerlingh Onnes Laboratory, Leiden University}
\author{Leonid Vidro}
\affiliation{Racah Institute of Physics, Hebrew University of Jerusalem, 91904 Jerusalem, Israel}
\author{Hagai S. Eisenberg}
\affiliation{Racah Institute of Physics, Hebrew University of Jerusalem, 91904 Jerusalem, Israel}
\author{Sophia E. Economou}
\affiliation{Department of Physics, Virginia Tech, Blacksburg, Virginia 24061, USA}

\begin{abstract}

Since linear-optical two-photon gates are inherently probabilistic, measurement-based implementations are particularly well suited for photonic platforms: a large highly-entangled photonic resource state, called a graph state,  is consumed through measurements to perform a computation. The challenge is thus to produce these graph states. Several generation procedures, which use either interacting quantum emitters or efficient spin-photon interface, have been proposed to create these photonic graph states deterministically. Yet, these solutions are still out of reach experimentally since the state-of-the-art is the generation of a linear graph state. Here, we introduce near-deterministic solutions for the generation of graph states using the current quantum emitter capabilities. We propose hybridizing quantum-emitter-based graph state generation with all-photonic fusion gates to produce graph states of complex topology near-deterministically. Our results should pave the way towards the practical implementation of resource-efficient quantum information processing, including measurement-based quantum communication and quantum computing.
\end{abstract}
\maketitle

Realizing a universal quantum computer is one of the long-sought objectives of quantum information technologies.
Photonic quantum processors are one of the leading contenders for the implementation of such a quantum computer as they are one of the only two platforms that have already reached the milestone of demonstrating quantum advantage~\cite{Zhong2020, Zhong2021}, together with superconducting circuits~\cite{Arute2019}.

The major challenge of a quantum processor based solely on linear optics is that photons do not interact with each other, and thus two-photon gates can only be implemented probabilistically through post-selection~\cite{Knill2001}.
Measurement-based quantum computing (MBQC) is potentially the way around this problem since quantum computation is only performed through single-qubit measurements on highly-entangled photonic graph states. Therefore, the major challenge of MBQC lies in the generation of a particular class of these highly-entangled photonic states, called graph states.
It was shown that the so-called 2D and 3D cluster graph states are respectively sufficient for universal~\cite{Raussendorf2001} and fault-tolerant~\cite{Raussendorf2006} quantum computation. Another important application of these graph states is for all-photonic quantum repeaters~\cite{Azuma2015, Ewert2016, Lee2019b, Hilaire2021b}, which are competitive with other quantum repeater schemes~\cite{Hilaire2021} when produced deterministically from a few matter qubits~\cite{Buterakos2017, Chan2018, Russo2018, Zhan2020}.

One approach for the generation of graph states is by using single-photon sources combined with linear optics to implement heralded probabilistic photon gates, fusion gates~\cite{Browne2005, Rudolph2017, Bartolucci2021}. This strategy relies on the heralded generation of 3-photon Greenberger-Horne-Zeilinger (GHZ) states that are fused to produce larger and larger graph states.
Since this strategy is heralded, the resource overhead can be reduced through the recycling of the graph states when a fusion gate has failed.
The practical implementation of this strategy requires a high single-photon detection probability~\cite{Varnava2008} and an adaptive platform based on the probabilistic fusion gate outcomes, which remains to be demonstrated experimentally for intermediate graph sizes. While promising for the long-term generation of a universal fault-tolerant quantum computer~\cite{Bartolucci2021},  this strategy may not be the most efficient for the short-term generation of graph states useful for the so-called noisy intermediate-scale quantum (NISQ) applications~\cite{Greganti2021}, such as photonic variational quantum eigensolvers ~\cite{Peruzzo2014, Ferguson2021}.

An alternative strategy, suitable for both NISQ and fault-tolerant quantum computing applications, is based on the deterministic generation of photonic graph states based on a few atoms or artificial atoms~\cite{Schon2005, Lindner2009}.
Several proposals follow this idea to produce photonic graph states of increasing dimensions and/or topological complexity. They are based either on small registers of interacting quantum emitters~\cite{Economou2010, Russo2019, Gimeno2019, Michaels2021, Li2022} or on a single quantum emitter with strong light-matter interaction~\cite{Pichler2017, Wan2020, Shi2021}.

Despite the large number of theoretical proposals, photonic graph states have only been generated probabilistically through passive linear optic generation~\cite{Zhong2018, Istrati2020, Zhang2022}, or deterministically using one single quantum emitter~\cite{Schwartz2016, Besse2020}. Ref.~\cite{Schwartz2016} constitutes the first demonstration of a potentially deterministic linear cluster state generation based on the Lindner and Rudolph scheme~\cite{Lindner2009}: it has experimentally demonstrated a 3-qubit linear cluster state through the manipulation of the dark exciton state of a quantum dot.
Even though this experiment was limited by the collection efficiency of the photons, it has also shown that the entanglement persists over 5 qubits using this linear cluster state generation, a number which has been recently improved up to 10 qubits~\cite{Cogan2021} using the same type of quantum emitters.
By improving the photon collection efficiency and using a single $^{87}$Rb atom, Ref.~\cite{Thomas2022} have recently demonstrated the generation and \emph{collection} of a $12$-photon linear cluster state and a $14$-photon GHZ state which constitute the largest entangled photonic state ever reported.

This discrepancy between the theoretical proposals and the experimental achievements lies in the fact that most of these graph state generation schemes rely on operations on the quantum emitters that are not simultaneously available for a given quantum emitter platform.
Creating arbitrary graph states with multiple atoms thus requires to have a platform which has both excellent photon emission properties, excellent atom control, and high fidelity atom-atom CZ gates. Yet, the current state-of-the-art quantum emitters have not met simultaneously these three criteria.
For example, arguably the best single-photon sources to date are based on quantum dots~\cite{Senellart2017}, which have a spin degree of freedom that we can use as an atomic qubit. However, these quantum emitters are generally not coupled to another spin register that could help produce graph states of larger dimensions\footnote{While Ref.~\cite{Jackson2021} showed that we can potentially use magnons as  additional quantum memories in quantum dots, it may remain unlikely that we can independently address more than one magnon simultaneously with high fidelities, given the experimental challenges of manipulating them.}. On the contrary, such a nuclear spin register exists in NV centers~\cite{Reiserer2016}, but the emitting properties of these defects are not as efficient as for quantum dots.

In this paper, we propose and evaluate ways to produce arbitrary graph states near-deterministically (for negligible photon losses) using only a single quantum emitter with a single spin degree of freedom and linear optics.
The two building blocks used in this scheme  --- emitter-based photonic graph state generation and boosted fusion gates --- are interesting in themselves and may also be used in other configurations or for other applications.
Besides, the boosted fusion gates that we introduce in this paper, can be made near-deterministic and uses much fewer resources compared to the already-existing ancilla-assisted schemes.

The paper is organized as follows.
In Section~\ref{sec_extended_linear_cluster}, we introduce a class of photonic states that can be realized with only a single spin quantum emitter. This class of photonic states are useful ``building blocks'' for the generation of multi-dimensional graph states. In Section~\ref{sec_hybrid_gen}, we introduce and analyze a boosted fusion gate which stands on its own but is also straightforward to use for the class of graph states presented in Section~\ref{sec_extended_linear_cluster}. We show that using these boosted fusion gates already leads to a clear success rate increase compared to type II fusion gates as long as the total loss per photon is less than $\approx 18 \%$. We then show in Section~\ref{sec_examples} how to produce larger and useful graph states near-deterministically using linear optics and deterministic single-spin graph-state generation, compatible with the current technological capabilities.

\begin{figure*}[!ht]
  \centering
  \includegraphics[width=2.1\columnwidth]{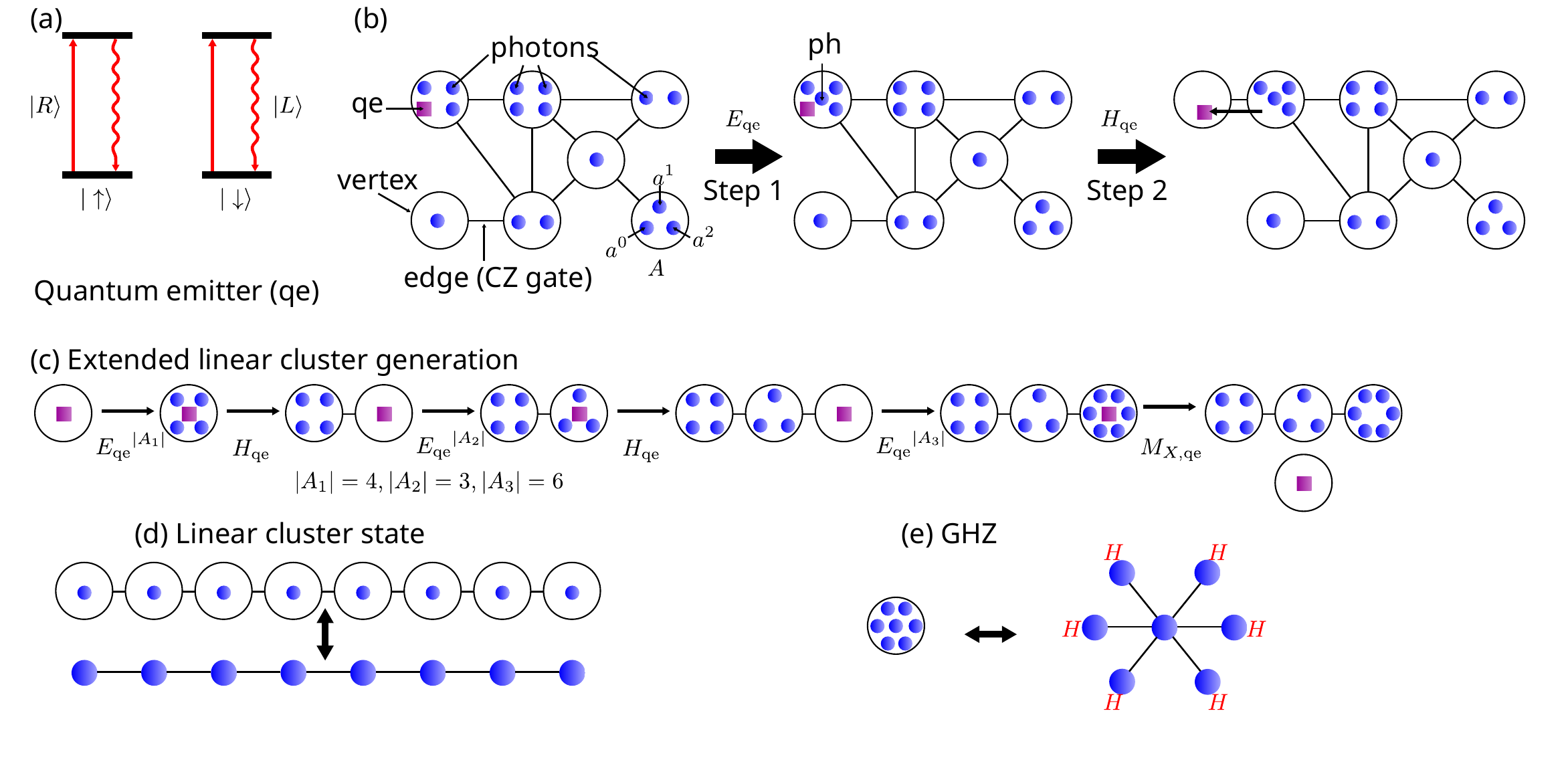}
  \caption{(a) Quantum emitter level structure. (b) Redundantly-encoded graph state and its general properties. Circles correspond to vertices, and points within circles correspond to qubits (blue for photonic qubits and purple for quantum emitter). Note that the Hadamard gate property would operate the same way on photons. (c) Redundantly-encoded linear cluster state generation using only one quantum emitter. It is illustrated with a three-vertex linear cluster (with $|A_1|=4,|A_2|=3,| A_3|=6$), but can be straightforwardly generalized to arbitrary  redundantly-encoded linear cluster. (d) Linear cluster state and (e) GHZ are represented as an  redundantly-encoded graph state and a usual graph state. (The GHZ state is local-unitary equivalent to a star graph and the local $H$ gate are also represented)}
  \label{fig_extended_graph}
\end{figure*}

\section{Redundantly-encoded linear cluster state generation}
\label{sec_extended_linear_cluster}

We start by evaluating a class of stabilizer states~\cite{Gottesman1997, Nielsen2002} that can be produced by a single emitter with a single qubit degree of freedom, such as an electron spin. We suppose that we can initialize and arbitrarily rotate the spin of the quantum emitter, which has a level structure shown in Fig.~\ref{fig_extended_graph}(a). Even though this level structure is convenient for the emission of a photon whose polarization is entangled with the spin degree of freedom, quantum emitters with different level structures can also be used to emit spin-entangled photons using different strategies (see for example Refs.~\cite{Lee2019, Tiurev2020, Tiurev2021}, where the photonic qubit is encoded in time bins). All the results from our work can be straightforwardly transferred to those quantum emitters.

\subsection{Redundantly-encoded graph states}

A graph state $\ket{G}$ is a quantum stabilizer state~\cite{Gottesman1997, Nielsen2002} that can be represented by a graph $G=(E,V)$ defined by an isomorphism in which the set of vertices $V$ correspond to qubits initialized in the $\ket{+} = (\ket{0} + \ket{1}) / \sqrt{2}$ state and the set of edges $E$ correspond to controlled-Z ($CZ$) gates between the pairs of qubits connected by each edge:
\begin{equation}
  \ket{G} = \left(\prod_{ (a,b) \in E} CZ_{a,b} \right) \mathop{\otimes}\limits_{a \in V} \ket{+}_a,
  \label{eq_graph_def}
\end{equation}
where $CZ_{a,b}$ corresponds to a $CZ$ gate between qubits $a$ and $b$.

We now propose a small extension of the class of graph states, referred to as ``redundantly-encoded graph states'', which are also stabilizer states and that will help with the discussion in the remainder of this paper.
In particular, we will show how a single quantum emitter can produce deterministically any redundantly-encoded \emph{linear} cluster state, from which general 2D/3D graph states can be easily produced.  For a redundantly-encoded graph state, we encode each qubit $a \in V$ of a graph with $n$ physical qubits using a repetition code ($n$ can be different for each logical qubit). Formally, we use a vector (denoted hereafter by a capital letter) $A = (a^0, a^1, ..., a^{n-1})$, where the $a^j$ correspond to the labels of the $n$ physical qubits that constitute the vertex {$A$}. The repetition code is such that the logical $\ket{0}_{A}$ (respectively $\ket{1}_{A}$) state is encoded with all its physical qubits in the $\ket{0}_{a^j}$ (respectively $\ket{1}_{a^j}$) state.

At the logical level, the redundantly-encoded graph state uses a similar definition   as Eq.~\eqref{eq_graph_def}, and gives at the physical level:
\begin{equation}
  \ket{G} = \left(\prod_{ (A, B) \in E} CZ_{a^{0},b^{0}} \right) \mathop{\otimes}\limits_{A \in V} \ket{+}_{A},
  \label{eq_ext_graph_def}
\end{equation}
with
\begin{equation}
  \ket{+}_{A} = \ket{\text{GHZ}}_{A} =  \frac{1}{\sqrt{2}}
   \left(
   \mathop{\otimes}\limits_{a^j \in A} \ket{0}_{a^{j}} +
   \mathop{\otimes}\limits_{a^j \in A} \ket{1}_{a^{j}}
   \right).
\end{equation}
and $CZ_{a^0,b^0}$ is a controlled Z gate between the physical qubits $a^0$, and $b^0$, which can be chosen arbitrarily from within the logical qubit since it is symmetric with respect to its physical qubits.
In the following subsection, we present properties and examples of these  redundantly-encoded graph states as in Fig.~\ref{fig_extended_graph}(b - e).

\subsection{Useful properties}

From Eq.~\eqref{eq_ext_graph_def}, it follows that if we perform a controlled-not ($CNOT$) gate between a physical qubit $a^j \in A$ ( $A \in V$) in the graph state and an isolated qubit initialized in the $\ket{0}$ state, this will position the new qubit at the same vertex  $A$ in the graph.
Interestingly, the emission of a spin-entangled photon by the quantum emitter, $E_{\rm qe}$, gives a similar result~\cite{Lindner2009}. The newly emitted photon is also positioned in the same vertex $A$ as the quantum emitter. Therefore, the $E_{\rm qe}$ operation is equivalent to ``initializing the photon qubit in the $\ket{0}_{\rm ph}$ state and applying a $CNOT$ gate between it and the spin qubit'':
\begin{equation}
  \begin{aligned}
    & E_{\rm qe} \left(\alpha \ket{H_0}\ket{0}_{\rm qe} + \beta \ket{H_1}\ket{1}_{\rm qe}\right) \\
    & = \left(\alpha \ket{H_0}\ket{0}_{\rm qe}\ket{0}_{\rm ph} + \beta \ket{H_1}\ket{1}_{\rm qe}\ket{1}_{\rm ph}\right),
  \end{aligned}
\end{equation}
where $\ket{H_0}$ and $\ket{H_1}$ are random quantum state acting on another set of qubits.
Therefore, as illustrated in Fig.~\ref{fig_extended_graph}(b)~step 1, we have the following property:
\begin{theorem}\label{prop_eqe}
  In a redundantly-encoded graph state, the emission of an entangled photon $E_{\rm qe}$, acts as adding a new physical photon qubit in the logical vertex where the quantum emitter is positioned.
\end{theorem}

This is not surprising since we can use a quantum emitter to produce GHZ photonic states of arbitrary size by successive emission of spin-entangled photons, which is exactly one isolated vertex.

\begin{figure*}[!ht]
  \centering
  \includegraphics[width=2.1\columnwidth]{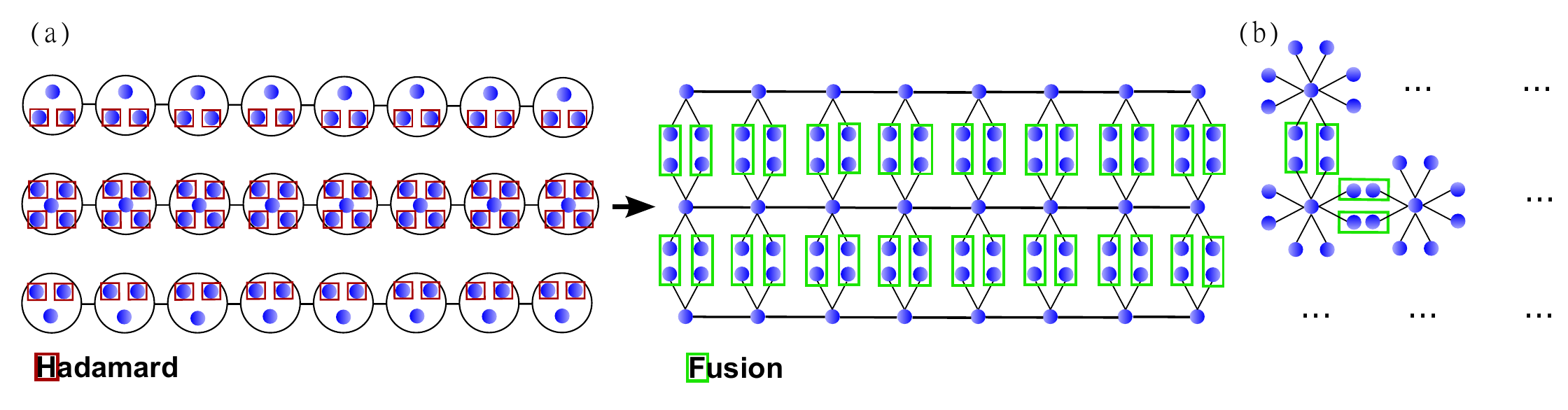}
  \caption{(a) Three layers of  redundantly-encoded linear cluster states. A Hadamard (marked in dashed red square) is applied on each but one of the physical qubits constructing a logical qubit, pushing them out and converting the state to a regular graph state. The resulting state resembles a linear cluster with ``dangling`` qubits that can used to connect to other layers via boosted fusion operations (marked by green dashed rectangles) (b).GHZ states which are simple  redundantly-encoded linear cluster states with pushed out qubits can also serve as building blocks for graph states by creating boosted links in two (or more) dimensions.}
  \label{pushed_out_examples}
\end{figure*}

Toward the manipulation of these redundantly-encoded cluster states, as illustrated in Fig.~\ref{fig_extended_graph}(b), a second important property of graph states is:
\begin{theorem}\label{prop_h}
  In a  redundantly-encoded graph state, if we apply a Hadamard gate ($H$) to a physical qubit $a^j$ with $|A| \geq 2$, this qubit is removed from vertex $A$ and is now in a new vertex connected by an edge to $A$.
\end{theorem}
From that, it becomes clear that a  redundantly-encoded graph state is local-Clifford equivalent to an actual graph state where the extra qubits at each vertex are "pushed out" from the vertices by applying $H$ gates. Fig.~\ref{fig_extended_graph}(e) shows an $n$-qubit GHZ state, which is also a logical qubit using a repetition code. It is equivalent to a star-shaped graph state if we apply a Hadamard gate on $n-1$ of these physical qubits. Indeed, such gates would push out these physical qubits out of the initial vertex to create a star topology. In Fig~\ref{pushed_out_examples}(a) three layers of  redundantly-encoded linear cluster states are converted to three layers of graph states, each resembling a linear cluster with ``dangling'' qubits that can be used to connect the layers to form a 2D cluster state. Fig~\ref{pushed_out_examples}(b) shows how star-shaped graph states (equivalent to GHZ states) can also be implemented to form more general cluster states. The ``dangling'' qubits can be used to connect the graphs in a straightforward way or by a boosted operation as will be explained in Section~\ref{sec_hybrid_gen}.

\subsection{Redundantly-encoded linear cluster generation}
\label{subsec_ext_lin_cluster_gen}

Thanks to these two previous properties, the generation scheme for a linear cluster state proposed by Lindner and Rudolph becomes clear~\cite{Lindner2009}. After preparing the spin in a $\ket{+}_{\rm qe}$ state, the emission process creates a photon at the same vertex (Prop.~\ref{prop_eqe}). Performing a spin $H$ gate positions the spin to a new vertex (Prop.~\ref{prop_h}), where a photon will then be emitted. Repeating successively the combination of the emission process and spin $H$ gates thus create a linear cluster state.

Besides, more general states than linear cluster states can be produced using a generalization of this idea, since  redundantly-encoded linear cluster states of arbitrary size can be produced with only one quantum emitter, as illustrated in Fig.~\ref{fig_extended_graph}(c).
A general  redundantly-encoded linear cluster, $\ket{G^{\rm (ext)}}$, with set of vertices  $V^{\rm (ext)} = \{ A_1, A_2, ..., A_N \}$ and with edge set $ E^{\rm (ext)} = \{(A_i, A_{i+1}) | i = 1...N-1\}$ can be produced using the following generation sequence:
\begin{equation}
  \ket{G^{\rm (ext)}} = M_{X, {\rm qe}} \left(\prod_{i = 1}^{N} {E_{\rm qe}}^{|A_i|} H_{\rm qe}\right) \ket{0}_{\rm qe},
  \label{eq_spin_gen_ext_lin_cluster}
\end{equation}
where ``qe'' denotes the quantum emitter qubit, $H_{\rm qe}$, $E_{\rm qe}$ and $M_{X, {\rm qe}}$, denote respectively a Hadamard gate, the emission of a spin-entangled photon, and the measurement in the $X$ basis of the quantum emitter. In this scheme, in order to prepare a new vertex a Hadamard is applied to the quantum emitter. Then, $|A_i|$ photons are emitted sequentially creating them in the same vertex. In the end, the quantum emitter is disconnected from the graph by an $X$ measurement

The linear cluster state is only a special case of a redundantly-encoded linear cluster state, where only one physical qubit is present at each vertex ($|A_i| = 1, \forall A_i \in V^{\rm (ext)}$). Its generation procedure is a particular case that can be straightforwardly deduced from Eq.~\eqref{eq_spin_gen_ext_lin_cluster}. Similarly, the GHZ generation procedure is simply a particular case of the trivial  redundantly-encoded graph with only one vertex ($V^{\rm (ext)} = \{ A_1\}$).
We illustrate these two special cases in Figs.~\ref{fig_extended_graph}(d), and (e).

Note that we can already see the potential of quantum-emitter-based generation to facilitate photonic quantum technologies.  We can use them to deterministically produce photonic GHZ states of arbitrary size using only one single quantum emitter. In comparison, the usual heralded linear-optical scheme generally requires at least 6 single-photon sources to produce a photonic GHZ$_3$ with $1/32$ success probability~\cite{Varnava2008}. More advanced adaptive schemes also exist and have better performances at the price of being more involved technologically, yet they remain inherently probabilistic~\cite{Bartolucci2021creation}.
Generating GHZ states efficiently is crucial for photonic quantum processing since they are a universal resource to produce larger graph states using fusion gates and linear optics. Yet, as we will see in the following, we can combine quantum emitters and linear-optic fusion gates to produce graph states more efficiently.

\section{Hybrid graph state generation}
\label{sec_hybrid_gen}

In general, it is not possible to create  redundantly-encoded graphs beyond this  redundantly-encoded linear cluster class (and local-Clifford equivalent states), without interaction between multiple quantum emitters or between a quantum emitter and photons~\cite{Li2022}. For example, the four-qubit square graph requires at least two interacting qubits.

Our strategy to overcome this limitation without increasing quantum resources (number of emitters) is to use probabilistic fusion gates between photons. We first review the type I and type II  heralded fusion gates~\cite{Pan2012, Browne2005}.
These two fusion gates have a success probability of at best 50\%, but we introduce new boosted type I and II fusion gates, with higher success probability, that are particularly convenient when dealing with  redundantly-encoded linear cluster states. We show that these boosted fusion gates have a better success probability, compared to other ancilla-assisted schemes by making use of the entanglement already present in the  redundantly-encoded linear cluster states.

\begin{figure}[!ht]
  \centering
  \includegraphics[width=\columnwidth]{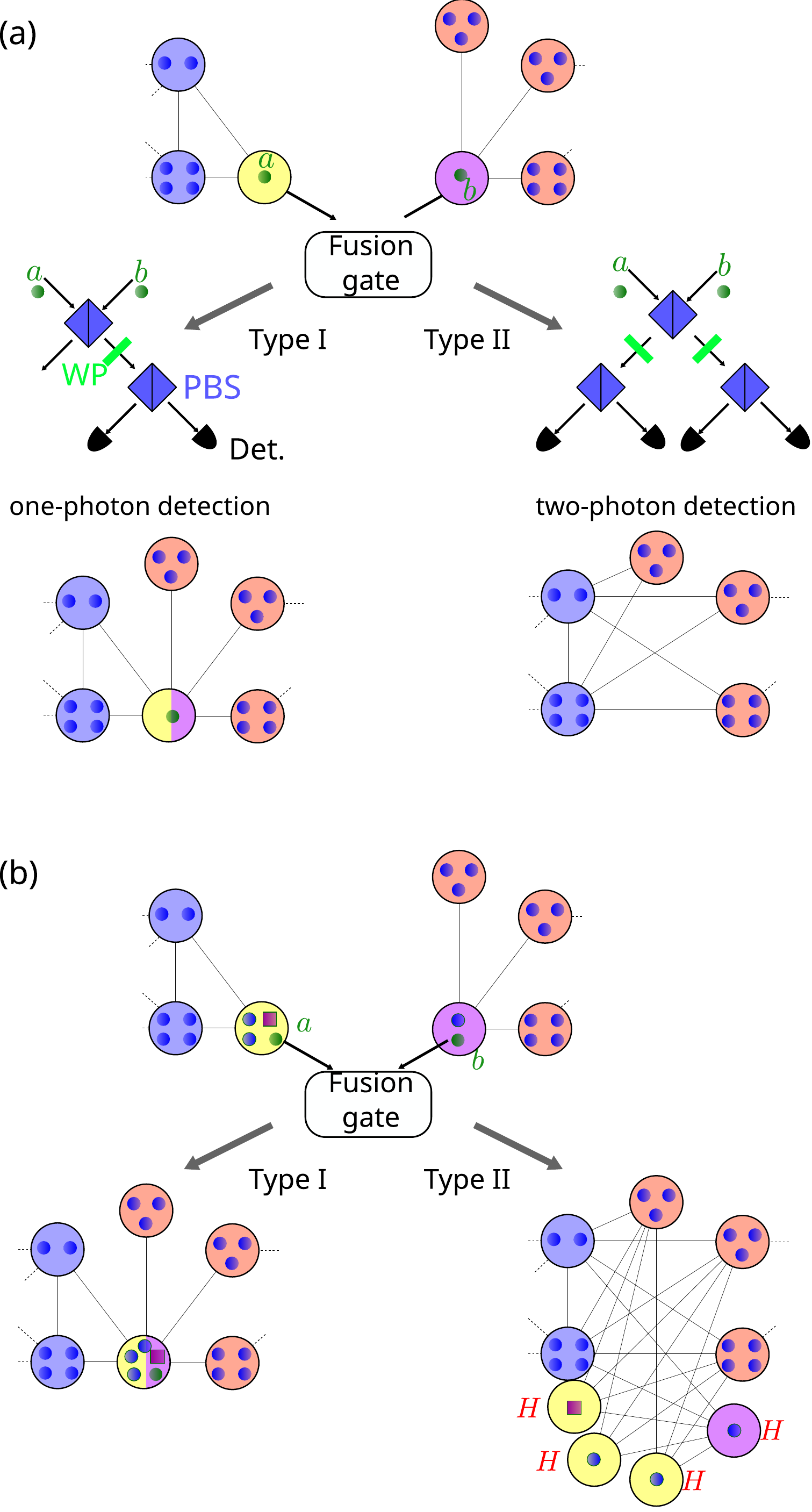}
  \caption{(a)Top: Fusion gate applied to photon qubits embedded in two  redundantly-encoded graph states. Each of the fused photons is the only qubit in its vertex. Middle: Linear optic setup for the different fusion gates. Bottom: Resulting graph state after application of fusion gates.  (b) An example where there are several physical qubits in each vertex of the fused qubits. For the type II fusion gate, the local $H$ gate that should be applied are also shown. PBS: polarizing beamsplitter, WP: waveplate, Det: detector.}
  \label{fig_fusion_0_I_II}
\end{figure}

\subsection{Type I and type II fusion gates in  redundantly-encoded graph states}

Type I and type II fusion gates are probabilistic operations that can merge two graphs. They are based on the detection of one photon for type I fusion gates and two photons for type II fusion gates and can be realized using the optical setups detailed in Fig.~\ref{fig_fusion_0_I_II}.
Type II fusion gates are more practical because photon loss is heralded: since two photons enter the optical setup and two photons should be detected to produce a successful gate, the loss of a photon cannot yield a successful result and thus can be handled accordingly.
The successful fusion gate corresponds to the successful detection of photons by the correct set of detectors (which occurs with probability at best 50 \% in the absence of photon loss). If the two photons trigger another set of detectors, the fusion gate fails but the photons are measured in the $Z$ basis, which effectively disconnects them from the graph. In that case, we say that the fusion gate \emph{partially fails}. In the presence of losses, if at least one photon is not detected, the fusion gate \emph{completely fails} and we don't know which output state is produced.

The behavior of both these fusion gates is well known for standard graph states (see for example Ref.~\cite{Browne2005}) and we need to generalize it to  redundantly-encoded graph states. Note that if we fuse qubits $a$ and $b$ which are the only qubits in their vertex (denoted $A$ and $B$ respectively), we can treat the redundantly-encoded graph case analogously to a normal graph state case (see Fig.~\ref{fig_fusion_0_I_II}(a)).
If there are other qubits at vertices $A$ and/or $B$ (see Fig.~\ref{fig_fusion_0_I_II}(b)), let's call these qubits $c^i$, the resulting  redundantly-encoded graph state is different. For \emph{type I fusion}, the resulting state is a fusion of $A$ and $B$, which contains all the qubits $c^i$ plus the qubit that was not measured after the fusion of $a$ and $b$. Indeed, we can easily predict this result  by looking at the local equivalent regular graph, by making the fusion gate and then by going back to the local equivalent  redundantly-encoded graph. By applying $H$ gates on all the $c^i$ qubits, we push all the $c^i$ qubits out of the vertex (Prop.~\ref{prop_h}), we then apply a type I fusion gate on regular graphs. We obtain the final expected result by applying $H$ gates again on the $c^i$ qubits to push them into the fused vertex. Note that we use these $H$ gates as an intermediate mathematical tool, but we don't need to actually apply them since they all commute with the fusion gate applied on qubits $a$ and $b$, and $H^2 = I$.

Using the same procedure, we can predict the effect of a \emph{type II fusion gate}, which is slightly more complicated, but we can show that the result is local unitary equivalent (up to $H$ gates on the $c^i$ qubits) to the following  redundantly-encoded graph state obtained using the following rule:
\begin{itemize}
  \item  Each neighbor of $A$ is now connected to each neighbor of $B$.
  \item Each $c^i$ qubit is now in its own vertex.
  \item If $c^i$ used to be in $A$ (respectively $B$), it is now connected to all neigbors  of $B$  (resp. $A$) and to all the $c^{i'}$ that used to be in $B$  (resp. $A$).
\end{itemize}

The intrinsic success probability of type I and II fusion gates is 50\%, yet type II fusion gates can be made near-deterministic by using more complicated interferometric setups that use additional ancillary photons~\cite{Grice2011, Ewert2014, Olivo2018}. In that case, a successful result requires detecting all the photonic qubits and ancillary photons. Therefore, the improved efficiency of these ancillary-assisted schemes also depends sensitively on the availability of near-deterministic single-photon sources and detectors. More recently, ancilla-assisted strategies have been  extended as well to type I fusion gates too~\cite{Bartolucci2021creation}.

In the case of a failed fusion, the detection of the photons leads to a $Z$ measurement on each photon. In a redundantly-encoded graph state, a $Z$ measurement disconnects not only the measured physical qubit from the graph but also the other physical qubits situated in the same vertex and thus completely disentangles the full vertex from the graph. To avoid this issue, we can first apply an $H$ gate on the qubits to push them out of their initial vertex (Prop.~\ref{prop_h}). By doing so, a failed fusion gate would only $Z$-measure the pair of qubits, preserving the rest of their initial vertices.  
In the following, we propose a strategy based on this idea to create near-deterministic fusion gates.

\subsection{Boosted fusion gates}

\begin{figure*}[!ht]
  \centering
  \includegraphics[width=2.1\columnwidth]{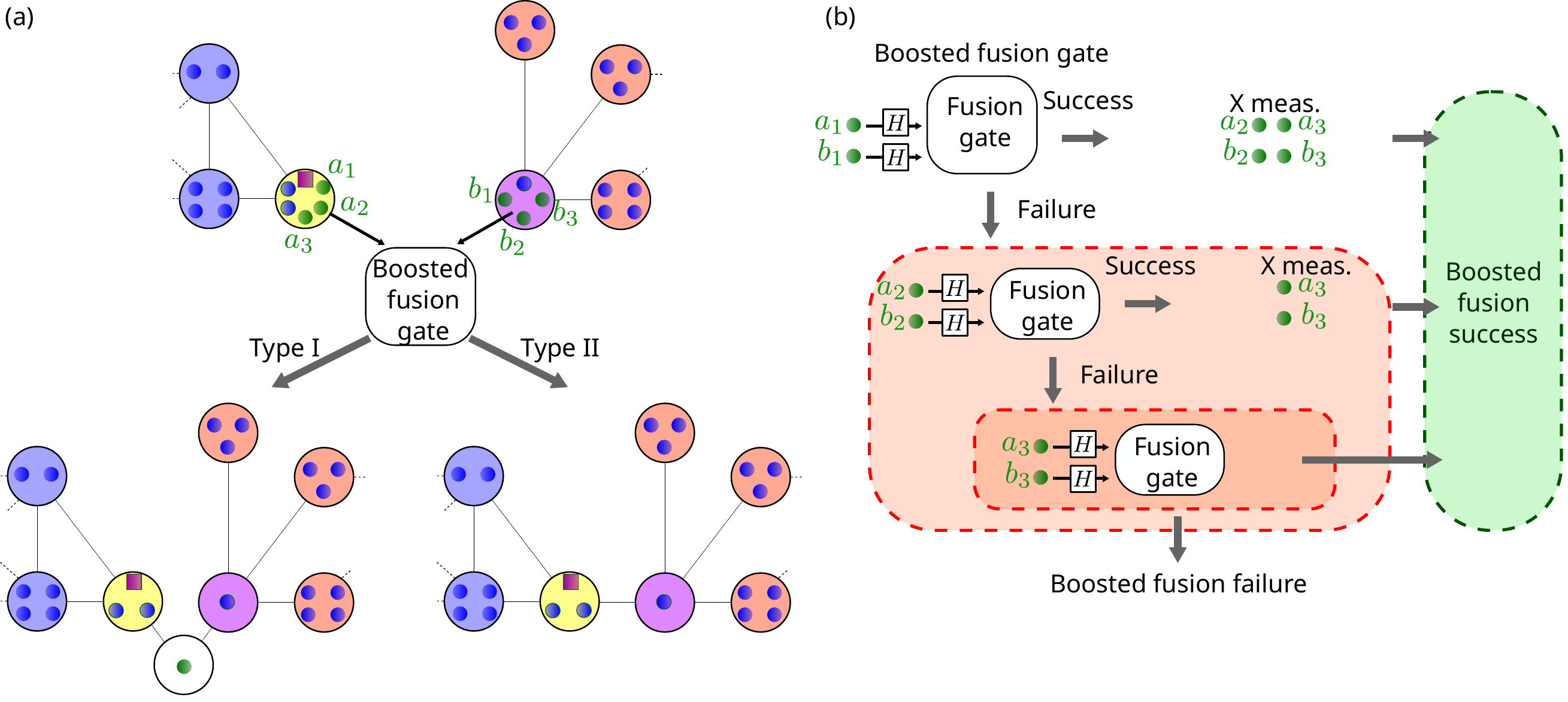}
  \caption{(a) Type I and type II boosted fusion gates effect on  redundantly-encoded graph states. (b) Measurement procedure for a boosted fusion gates. The procedure is the same for type I and type II boosted fusion gates. (The results depicted here are for applying $H$ gates to the fused qubits before fusion, which first pushes them out of their initial vertices).}
  \label{fig_boosted_fusion}
\end{figure*}

With ancillary-assisted methods~\cite{Grice2011, Ewert2014, Olivo2018}, fusion gates can be efficiently made near-deterministic, by using ancillary photons that are not part of the graphs. However, this strategy uses generally a large overhead to approach 100 \% success probability. Here, we propose a different strategy, which makes use of the possibility of producing  redundantly-encoded linear clusters efficiently. Therefore, we can use extra photons positioned at the same vertex rather than ancillary photons.
We will show that such a boosted fusion gate reaches the fundamental maximum success rate as derived in Ref.~\cite{Lee2019b} and is thus actually much more efficient than using ancillary qubits.

We illustrate the general idea behind boosted fusion gates in Fig.~\ref{fig_boosted_fusion} and we detail it in the following.
The key idea of boosted fusion gates is that we allocate $m$ qubits per vertex for vertices $A$ and $B$, to fuse them ($m<\min ({|A|, |B|})$.
Note that if we want to fuse one vertex $|A|$ with $k$ other vertices, we will need $mk$ qubits for all these fusions and one final qubit that will remain in this vertex, hence $|A|\geq m k+1$.

We denote these qubits $a^1, ..., a^m$ and $b^1, ..., b^m$. We first apply $H$ gates on $a^1$ and $b^1$ (to ``push'' them out of the vertex) and apply a fusion gate onto them (either type I or II fusions). We use a variant of the type II fusion which is described in Appendix~\ref{app_fusion} instead of the usual fusion gate, and explain why we opt for this choice. For the moment, we consider the loss-less case where the complete failure of a fusion gate is impossible, and we will consider photon losses later. If the fusion gate succeeds, we measure the remaining qubits $a^2, ..., a^m$ and  $b^2, ..., b^m$ in the $X$ basis to remove them from vertices $A$ and $B$ and we end up with an entanglement connection between $A$ and $B$. Alternatively, if other connections are planned for those vertices, the remaining qubits can serve to boost those as well. Yet in the following, for simplicity, we will only allocate each qubit for a specific boosted fusion.
Now, we will examine what happens if a partially failed fusion gate occurs.
Since a failed fusion gate corresponds to two individual $Z$ measurements, the qubits are automatically disconnected from the  redundantly-encoded graph. We can therefore apply the same procedure to $a^2$ and $b^2$ and so on until a successful fusion gate occurs, or until we reach a failed fusion gate on $a^m$ and $b^m$. The boosted fusion gate fails only in that case.

Our boosted fusion scheme has strong similarities with the repeat-until-success (RUS) scheme from Ref.~\cite{Lim2005}, which allows to near-deterministically entangle distant quantum emitters using photonic qubits. In this scheme, each quantum emitter emits an entangled photon on which a fusion gate is applied. If the fusion is successful, it performs an entangling gate on the quantum emitters, and if unsuccessful, the process is repeated until it succeeds (or a photon is lost). Similarly to our boosted fusion gate, the RUS scheme is near-deterministic (assuming no photon losses) and is also not loss-tolerant. 

The boosted fusion gate scheme can be viewed as a ``time-delayed'' adaptation of the RUS scheme. Given its time-delayed nature, we need to allocate a given number of photon to a fusion, which is not the case for the RUS scheme. Because of this, our boosted fusion scheme performs slightly worse than the RUS scheme in terms of success probability (the interested reader can find a quantitative comparison with the RUS scheme and ancilla-assisted fusion gates in Appendix~\ref{sec_comp}).
However, the RUS scheme needs to allocate two matter qubits during the full time of the fusion which is not the case for a boosted fusion gate. This provides greater flexibility in the photonic graph state generation, which could be useful to reduce the matter qubit overhead to generate a graph state. For example, in the proposal of graph state generation~\cite{Barrett2005, Lim2006} using the RUS scheme, we need a matter qubit at each node of the graph state. In comparison, we will show that the number of resources needed to generate a given graph state is much smaller.

Because the repetition code is not fully quantum error-correcting, using boosted fusion gates leads to an increased sensitivity to errors. However, the boosted fusion gate error rate only grows linearly with $m$ compared to the usual fusion gates, while the failure probability is exponentially suppressed.
Indeed, assuming no photon losses, we obtain a near-deterministic fusion gate, with success probability $1 - 2^{-m}$. The failure probability of these boosted fusion gates therefore decreases exponentially with the amount $2m$ of photonic qubit used. This is in stark contrast with ancilla-assisted fusion gates where the number of ancilla photons grows exponentially. We emphasize that this success probability is the best achievable with linear optics when using $2m$ photons, as derived in Ref.~\cite{Lee2019b}. Compared to this result, our boosted fusion gate uses a simple repetition code, which has the drawback of not being loss-tolerant, but also the great advantage of being deterministically generated with quantum emitters.
In particular, it is well suited to the deterministic  redundantly-encoded linear cluster state generation discussed in Sec.~\ref{subsec_ext_lin_cluster_gen} since adding photonic qubits to a given vertex can be done efficiently. In the following, we focus our analysis on type II boosted fusion gates.

\begin{figure}[ht]
  \centering
  \includegraphics[width=\columnwidth]{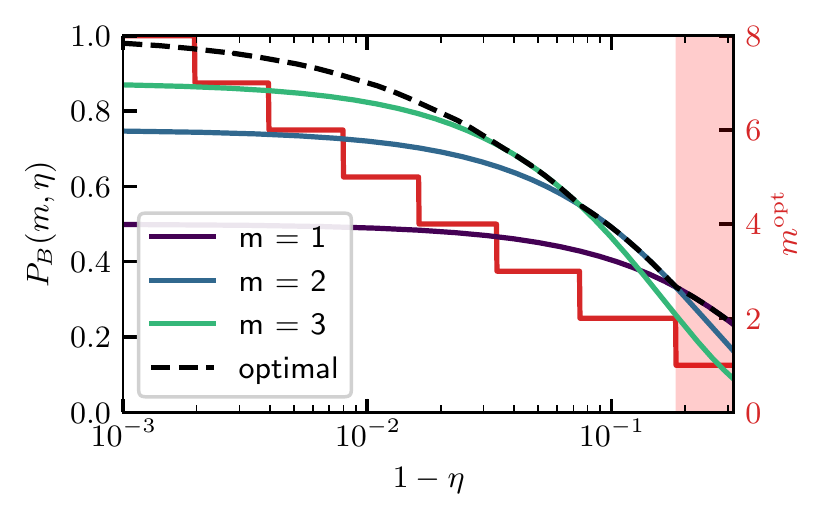}
  \caption{Success probability of the boosted fusion gate $P_B(m, \eta)$ as a function of the photon loss probability ($1-\eta$). The optimized number of photons $m^{\rm opt}$ (red curve) that maximizes the success probability $P_B(m^{\rm opt}, \eta)$ (dashed line) is also represented. The case $m=1$ corresponds to the usual type II fusion gate. The pink area corresponds to the region where boosted fusion gates perform equivalently as usual type II fusion gates ($\eta \leq \sqrt{2/3}$)}
  \label{fig_boosted_fusion_opt}
\end{figure}

In the presence of losses, the boosted fusion gates can only succeed if all the $2m$ photonic qubits involved are actually detected. Let's denote by $\eta$ the probability that a photonic qubit is detected. The success probability of a  type II boosted fusion gate is therefore
\begin{equation}
  P_B(m, \eta) = (1 - 2^{-m}) \eta^{2m}.
  \label{eq_p_boosted}
\end{equation}
Consequently, increasing the number $2m$ of qubits used for the fusion gates may decrease the success probability of the boosted fusion gate if $\eta \neq 1$. The optimal value of $m$, $m^{\rm opt}$, which maximizes the success probability of boosted fusion gates $P_B$ follows the condition for which the improvement factor $f(m)$ in success probability from adding an additional fused pair is greater than the reduction factor in overall detection efficiency of this pair, $\eta^2$:
\begin{equation}
  \begin{aligned}
    f(m^{\rm opt} - 1) &  \leq \eta^2 < f(m^{\rm opt}) \\
    {\rm with \;} f(m) & =
    \begin{cases}
      0, & \text{if } m = 0\\
    \frac{1 - 2^{-m}}{1 - 2^{-(m+1)}},
    & \text{otherwise.} \\
    \end{cases}
  \end{aligned}
  \label{eq_m_opt}
\end{equation}
The boosted fusion gates (both type I and II) provide an advantage if the overall detection efficiency is above $\eta > \sqrt{f(1)} = \sqrt{(2/3)} \approx 82 \%$. Below this value, we can still use type II fusion gates for the hybrid generation scheme, which should still provide an advantage over linear optics.
Fig.~\ref{fig_boosted_fusion_opt} illustrates the dependence between the optimized success probability of the boosted fusion gates and the single photon detection probability: for a given value of $\eta$, we choose $m^{\rm opt}$ to maximize the boosted fusion gate success probability, $P_B(m^{\rm opt},\eta)$.
The advantage of this scheme is significant in the low loss regime. Moreover, we can benefit from additional ideas presented for graph states under loss.
Suppose that we have a linear cluster of redundantly encoded logical qubits.
Since a logical qubit in this scheme resembles a GHZ state, loss of a physical qubit (one photon) causes the loss of the entire logical qubit. However, the loss tolerance on the logical level depends on the graph topology. For example, upon loss of a logical qubit $A$ on a linear redundantly-encoded graph state, two shorter linear clusters before and after vertex $A$ can be retrieved by measuring the neighbors of $A$ in the $Z$ basis. We can reuse these graphs for other purposes (see for example~\cite{Varnava2008, Pant2017}).
Furthermore, additional constructions can be used on top of the proposed encoding, such as \cite{Varnava2006,Bell2022} to cope with photon loss using quantum error correcting codes.

\section{Examples of graph state generation}
\label{sec_examples}
\begin{figure}
  \centering
  \includegraphics[width=\columnwidth]{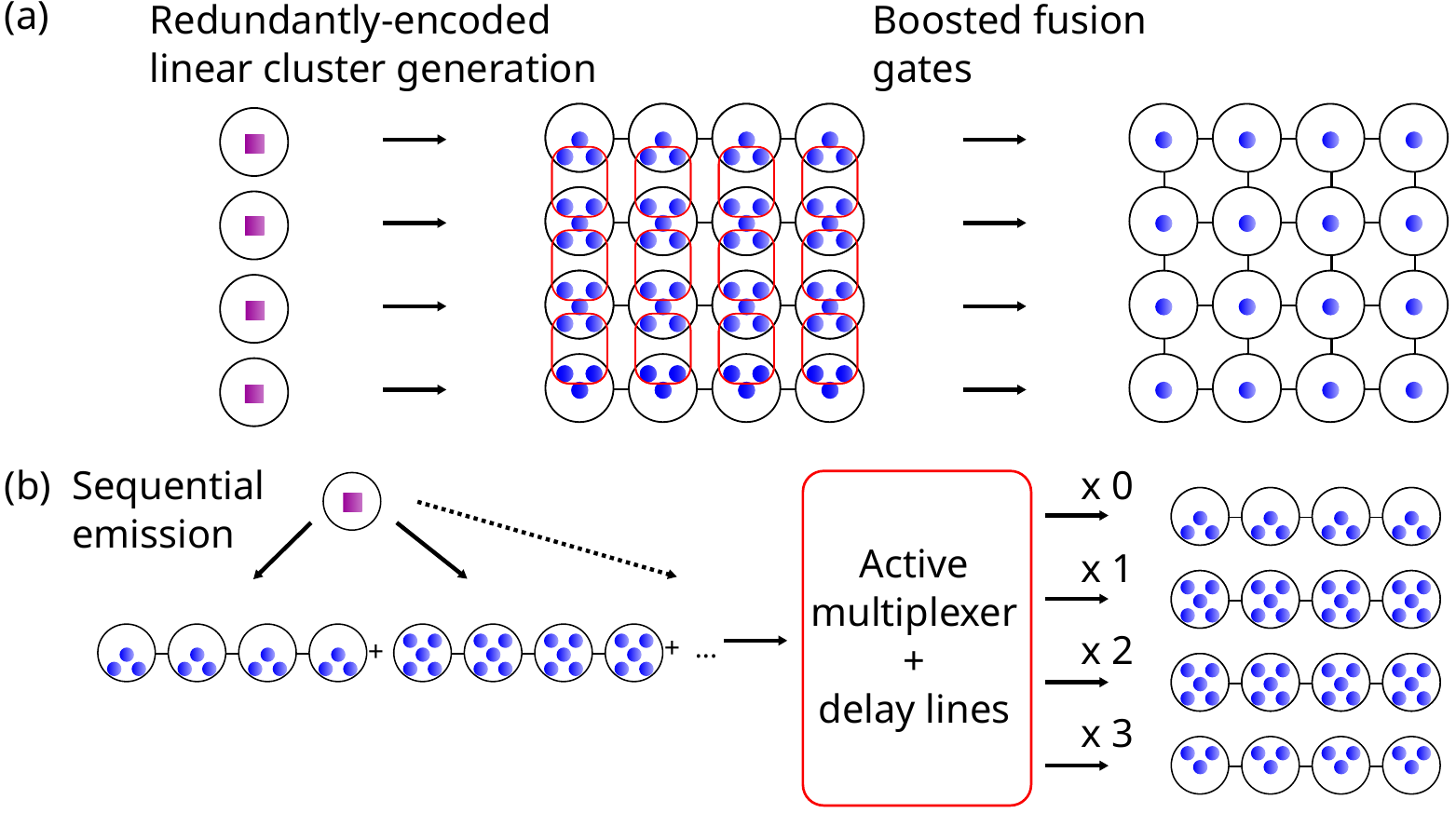}
  \caption{(a) Generation of a 2D cluster state using multiple quantum emitters in parallel and boosted fusion gate. (b) Modified generation scheme using one emitter to sequentially produce the  redundantly-encoded linear cluster states. It also requires an active multiplexer and delay lines (the number of delay lines per output mode of the multiplexer is also indicated)}
  \label{fig_seq}
\end{figure}

In the following, we propose generation schemes for two categories of graph states used in the measurement-based paradigm. However, arbitrary graph states can be created using those ideas with appropriate setups, so it should be possible to extend this work to other interesting graphs such as repeater graph states and others. The first family of graphs is multi-dimensional cluster states, which can be used for universal and fault-tolerant quantum computing. The second category of graphs are unencoded and encoded ring graph states which are the key resource for fusion-based quantum computing. As a first step, we discuss how a single quantum emitter with a spin-photon interface combined with linear optics is sufficient to produce these graph states.

\subsection{Multiple non-interacting quantum emitters vs one single quantum emitter}

In general, we can divide the generation scheme into two steps: first, we produce the  redundantly-encoded linear cluster states and then we use the linear-optic boosted fusion gates to entangle them.

Multiple quantum emitters can produce the  redundantly-encoded linear cluster states in parallel (see Fig.~\ref{fig_seq}(a)). This strategy requires many quantum emitters able to produce mutually indistinguishable photons since the fusion gates only work for perfectly identical photons. This is different and much more challenging than having indistinguishable photons produced from a single quantum emitter~\cite{Kambs2018}.
Many experimental works have probed either moderate indistinguishability between distinct single-photon sources~\cite{Beugnon2006, Maunz2007, Patel2010, Giesz2015, Gold2014, Bernien2012},
or rely on spectral filtering~\cite{Bernien2013, Sipahigil2014, Stockill2017, Delteil2016}  which intrinsically reduce the single-photon source brightness~\cite{Senellart2017}. Ideally, the challenge is therefore to have multiple controllable quantum emitters as sources of indistinguishable photons without having to trade off indistinguishability over brightness.

We also consider an alternative approach where we produce any graph state with a single quantum emitter. The idea is simply to emit the necessary reduntantly-encoded linear cluster state sequentially by a unique quantum emitter. With this strategy, illustrated in Fig.~\ref{fig_seq}(b), all the photons originate from the same quantum emitter, and thus, making them indistinguishable is less technologically demanding. Indeed, bright sources of indistinguishable photons have been already demonstrated with quantum dots~\cite{Somaschi2016, Ding2016, Uppu2020, Tomm2021, Cogan2021, Coste2023} and defects in diamond~\cite{Riedel2017, Zhang2018, Knall2022}.
Therefore, we can use equivalently either sequential emission from a single quantum emitter or parallel emission from multiple quantum emitters if they can emit mutually indistinguishable photons. For the sake of simplicity, we opt for the latter to describe the following graph state generation schemes.

\subsection{Hybrid 2D/ 3D / $d$-dimensional cluster state generation}

\begin{figure*}[ht]
  \centering
  \includegraphics[width=2.1\columnwidth]{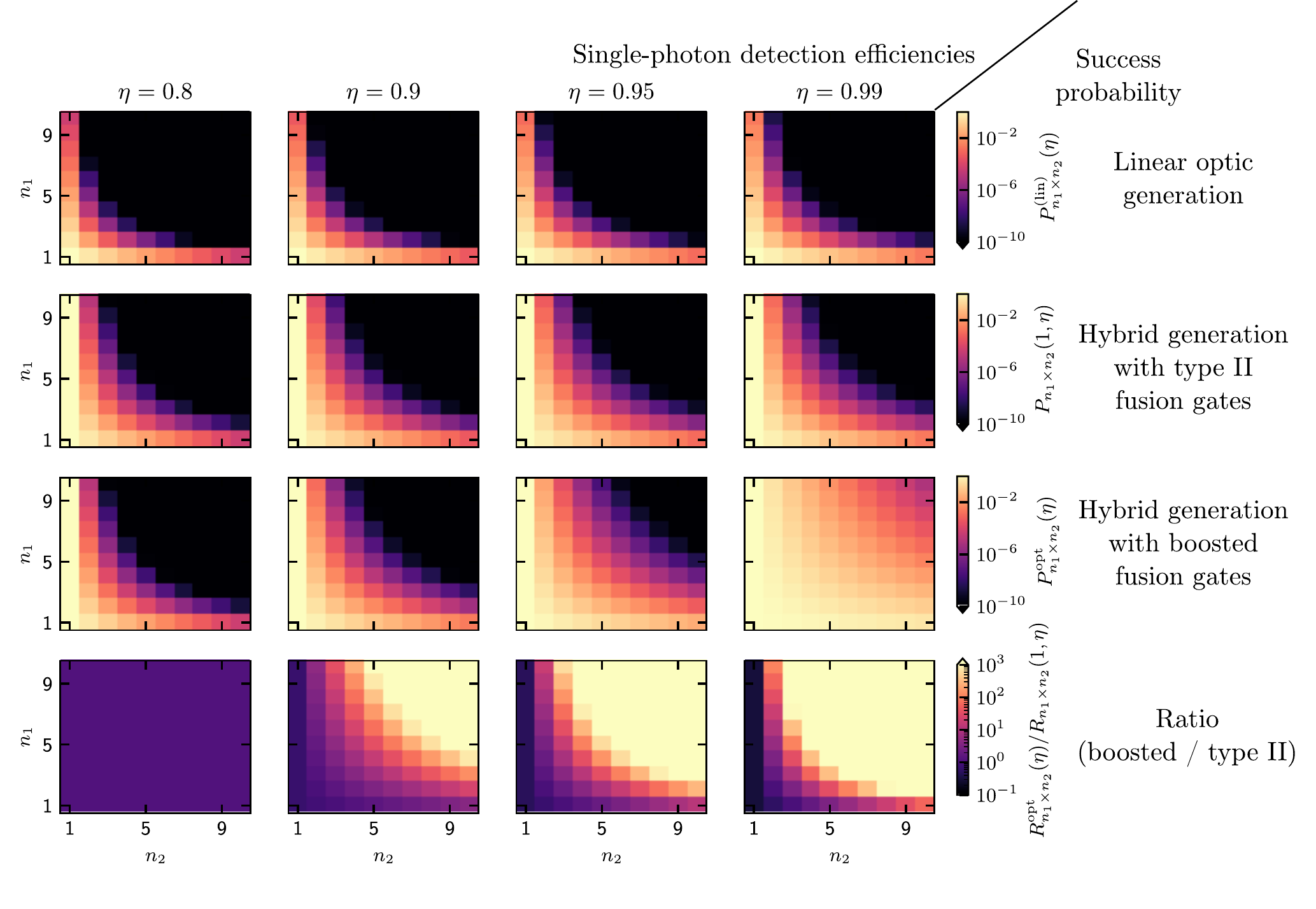}
  \caption{Success probability for the generation of $n_1 \times n_2$ cluster states for different single-photon detection probabilities $\eta$ (columns). The first row corresponds to an estimate of the generation probability using non-adaptive all-photonic scheme. The second (respectively third) row corresponds to a hybrid generation using non-interacting spins and type II (respectively  boosted) fusion gates. The fourth row shows the ratio between the generation rates, when using boosted fusion gates rather than type II fusion gates. Values below $10^{-10}$ (respectively above $10^3$) for the first three rows (resp. the fourth row) are not detailed.}
  \label{fig_cluster_perf}
\end{figure*}

In the following, we will only focus on 2D cluster states but the general ideas are applicable to multi-dimensional graph states as well, as we discuss in Appendix~\ref{sec_3D_cluster}.
A $n_1 \times n_2$ 2D cluster state can be produced as illustrated in Fig.~\ref{fig_seq}(a) and detailed in the following.
$n_2$ quantum emitters produce  redundantly-encoded linear cluster with $n_1$ vertices, each including $2 m + 1$ photons (except for the first and the last quantum emitters which should emit redundantly-encoded linear clusters that only include $m + 1$ photons), where $m$ is the number of photons per vertex consumed during each type II boosted fusion gate. The total number of boosted fusion gates is thus $(n_2-1)n_1$ and each has an individual success probability $P_B(m, \eta)$ (see Eq.~\eqref{eq_p_boosted}), so that the overall success probability is:
\begin{equation}
  P_{n_1\times n_2}(m,\eta) = P_B(m, \eta)^{n_1(n_2-1)}.
\end{equation}
Furthermore, we can choose $m = m^{\rm opt}$ using Eq.~\eqref{eq_m_opt}, to maximize the success probability of a graph state generation, $P_{n_1\times n_2}^{\rm opt}(\eta) = P_{n_1\times n_2}(m^{\rm opt},\eta)$.  Also note that in the case of negligible losses $\eta \to 1$, $m^{\rm opt} \to \infty$, but the success probability can be made arbitrarily close to 1 since $P_{n_1\times n_2}(m,1) = (1 - 2^{-m})^{n_1(n_2-1)} \to 1$ when $m \to \infty$.

The resulting graph state is thus a $n_1 \times n_2$ graph where each photon can be detected with probability $\eta$.
In Fig~\ref{fig_cluster_perf}, we compare the success probabilities of three different strategies to produce $n_1 \times n_2$ cluster states. We first show an estimate of the success probability for an all-photonic non-adaptive scheme where we assume that each edge is generated with success probability $\eta^2/ 2$. This clearly illustrates that all-photonic strategies are not efficient for intermediate scale graph state generation without adaptive schemes and feedforward. Then, we compare two hybrid generation schemes where we use deterministic  redundantly-encoded linear cluster generation and either type II fusion gates or optimized type II boosted fusion gate as discussed before. These results clearly illustrate the strong improvement of hybrid schemes over all-photonic generation. Besides, for any $\eta > \sqrt{(2/3)}$, the boosted fusion gates provide an advantage and this advantage is all the more significant when $\eta$ is close to $1$ and when the cluster state is large. While the probabilities may seem relatively small, it is important to note that quantum emitters emit photons at rates that can reach the 1-10GHz regime~\cite{Liu2018}, so that even having an overall generation probability of $P \approx 10^{-9}$ for a cluster state can lead to a generation rate of 10mHz-1Hz. For a fair comparison of these two generation strategies, we should divide this probability by the generation time of one  redundantly-encoded linear cluster state:

\begin{equation}
  T_{\rm ext}(m, n_1) = n_1((2m+1) T_{E_{\rm qe}} +  T_H),
\end{equation}
with $m = m^{\rm opt}$ for the boosted fusion gate strategy and $m=1$ for the type II fusion gates. Here $T_{E_{\rm qe}}$ and $T_H$ correspond to the times for the photon emission and a spin Hadamard gate respectively. The success rate is thus $R_{n_1\times n_2}(m,\eta) = P_{n_1\times n_2}(m,\eta) / T_{\rm ext}(m, n_1)$.
For example, we can see in Fig.~\ref{fig_cluster_perf} that the generation rates using boosted fusion gates can be much greater compared to using standard fusion gates. For example, with detection efficiencies of $\eta = 0.9$ and $0.95$, the generation rate of a $5 \times 5$ cluster state is improved by a factor $\approx 30$ and a factor $\approx 500$, respectively. For $\eta=0.95$, the success probability of the entire graph with the boosted fusion gate is typically of $\approx 10^{-4}$ which should be put in perspective with the high photon emission rates of usual quantum emitters (around the GHZ range).

Even assuming a negligible spin gate time compared to the emission process, i.e. a comparison which favors type II gates, we see a clear improvement of the success rate.
 It is therefore clear that our fusion scheme provides a practical improvement for detection probabilities above $\eta \geq 90\%$.

\subsection{Ring graph state generation}

  \begin{figure*}
    \centering
    \includegraphics[width=2.1\columnwidth]{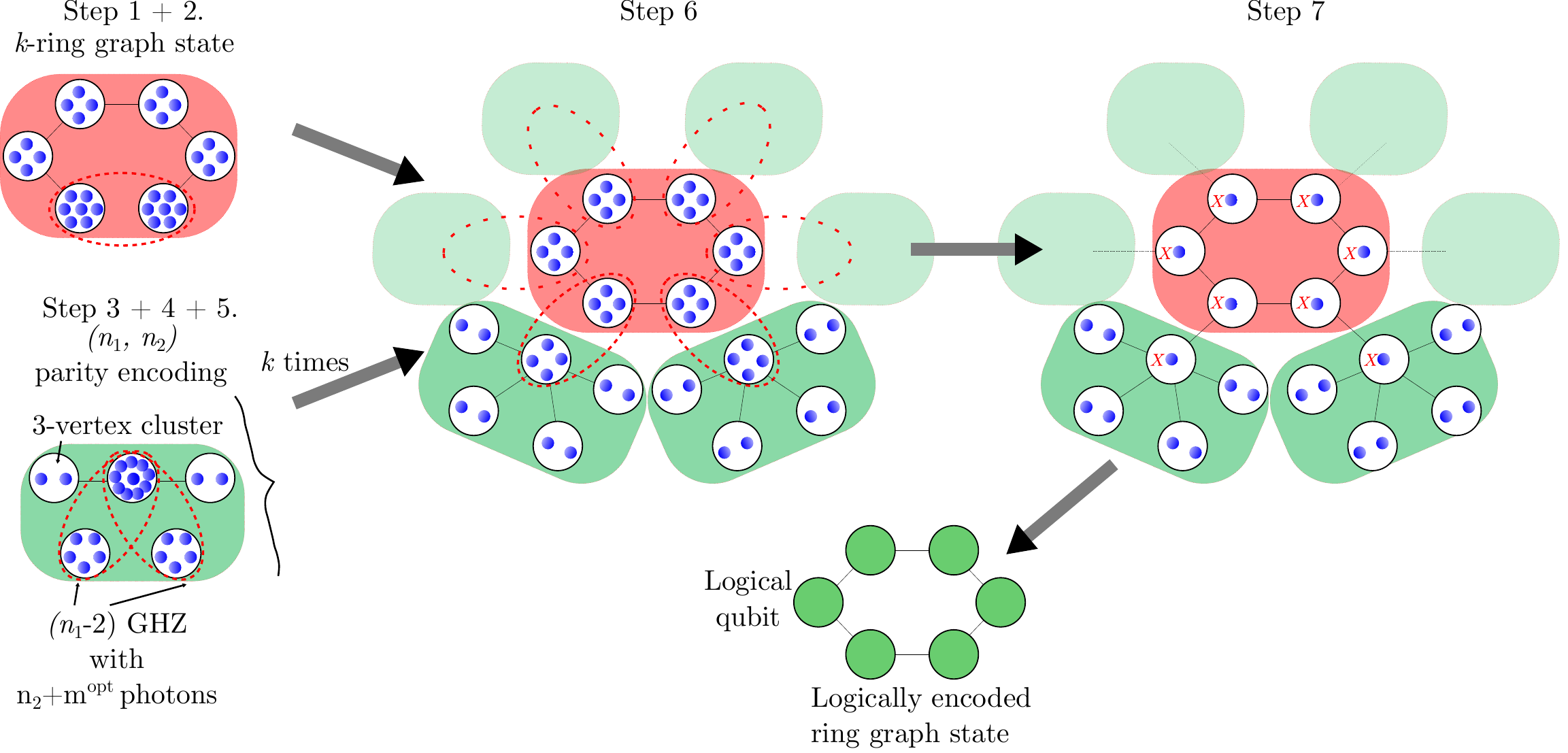}
    \caption{Generation procedure of encoded ring graph states using quantum emitters and linear optics. Step 1+2: Generation of an unencoded $k$-ring graph state (here with $k=6$). Step 3+4+5: Generation of a logical qubit, which uses a $(n_1, n_2)$ parity encoding (with $n_1=4$ and $n_2=2$). Step 6+7: Generation of the $C_{(k,n_1,n_2)}$ logically-encoded graph state through type II boosted fusion gates (step 6) and $X$ measurements (step 7). All the boosted fusion gates are represented by dashed red lines and uses $m^{\rm opt} = 3$ photons.}
    \label{fig_ring_graph}
  \end{figure*}

For fusion-based quantum computing~\cite{Bartolucci2021}, the resource states are either 4-qubit GHZ states, 6-qubit ring graph states or encoded 6-qubit ring graph states using quantum parity encoding. 4-qubit GHZ states can be produced deterministically but are highly sensitive to photon loss.

\paragraph{$k$-qubit ring graph states.}
6-qubit ring graph states can be produced by first deterministically generating a 6-vertex  redundantly-encoded linear cluster state, with $1$ photon at each vertex except the first and the last which each contains $1+m^{\rm opt}$ photons.  Then, we use a single type II boosted fusion gate on $m^{\rm opt}$ photons of the first and last vertices to produce the 6-qubit ring graph states. Producing a $k$-qubit ring graph state follows the same procedure and always succeeds as long as the single boosted fusion gate succeeds.

\paragraph{Encoded $k$-qubit ring graph states.}
The performance of FBQC, and in particular its tolerance to photon loss and gate failures, can be significantly increased by adding a quantum parity encoding to the ring graphs. In that case, the qubits are now logically encoded using a $(n_1, n_2)$ parity encoding~\cite{Ralph2005}. It is possible to realize this using our hybrid generation scheme.
The generation procedure of a $k$-qubit ring which uses a $(n_1, n_2)$ parity encoding, denoted $C_{(k,n_1,n_2)}$, is the following (see also Fig.~\ref{fig_ring_graph}):

Creation of a $k$-qubit ring graph state:
\begin{enumerate}
  \item First, produce a  redundantly-encoded linear cluster with $m^{\rm opt} + 1$ photons at each vertex except the first and last vertex which contains $2m^{\rm opt} + 1$ photons.
  \item Perform a single type II boosted fusion gate between the first and last vertices of the  redundantly-encoded linear cluster to obtain a  redundantly-encoded ring state with $m^{\rm opt} + 1$ photons per vertex. Compared to the unencoded $k$-qubit ring graph state, the additional $m^{\rm opt}$ photons per vertex will be used to attach the logical qubits (see step 6).
\end{enumerate}
  \noindent\rule{7.5cm}{0.4pt}
Creation of the $k$ copies of ($n_1, n_2$) parity-encoded logical qubits:
\begin{enumerate}[resume]
  \item Produce $k$ 3-vertex  redundantly-encoded linear cluster with $n_2$ qubits in the first and last vertices, and $(n_1-1) m^{\rm opt} + 1$ qubits in the middle one. One set of $m^{\rm opt}$ photons of the middle vertex will be used to fix the logical qubit to the ring (see step 6). $(n_1 - 2)m^{\rm opt}$ photons will be used to complete the logical qubit (see step 5). The remaining photon will be $X$-measured in step 7.
  \item Produce $k \times (n_1 - 2)$ GHZ states of $(n_2 + m^{\rm opt})$ photons.
  \item Perform a type II boosted fusion gate between the middle vertex of each of the 3-vertex  redundantly-encoded linear clusters and $n_1-2$ of the GHZ states (each GHZ state only fuses with one of the $k$ redundantly encoded 3-vertex cluster states). This operation creates a parity encoded logical qubit (with one additional vertex), which is a redundantly-encoded star graph of $n_1 + 1$ vertices. This has to be repeated $k$ times, to prepare the $k$ logical qubits.
\end{enumerate}
    \noindent\rule{7.5cm}{0.4pt}
Creation of the encoded $k$-qubit ring ($C_{(k,n_1,n_2)}$):
\begin{enumerate}[resume]
  \item Perform $k$ type II boosted fusion gates between the middle vertex of each star graph and a vertex of the  redundantly-encoded $k$-qubit ring state.
  \item $X$-measure the $k$ qubits that were initially in the $k$-qubit ring graph, and the $k$ photons that were originally in each middle vertex of the  redundantly-encoded star graphs.
\end{enumerate}
The generation procedure is heralded, i.e. we know when the encoded ring graph has been generated and does not require feedforward. Therefore, the success probability is calculated assuming that each of its steps has succeeded.
The total number of type II boosted fusion gates is $1 + k(n_1 - 1)$ and the overall success probability of this scheme is
\begin{equation}\label{eq_p_c_k_n}
 P_{C_{(k,n_1,n_2)}}^{\rm opt}(\eta) = P_B(m^{\rm opt}, \eta)^{1 + k(n_1 - 1)} \eta^{2k},
\end{equation}
Note that this expression does not depend on the number of photons per block, $n_2$, of the parity encoding, which are produced deterministically thanks to the quantum emitters.

An encoded ring graph state requires significantly more fusion gates than a simple ring graph state. This is the price to pay to reach higher loss tolerance. In practice, for fusion-based quantum computing, we need to produce these encoded (or unencoded) ring graph states with high success probability, thus requiring to parallelize the generation to increase the chance of successfully creating one. Using $N$ generation schemes in parallel, we produce at least one encoded ring graph state with a success probability $1 - (1 - P_{C_{(k, n_1, n_2}}(\eta))^N$. In practice, $P_{C_{(k,n_1,n_2)}}^{\rm opt}(\eta)$ is expected to be very small, in particular for moderate or large values of $k$, $n_1$, and $n_2$, thus requiring massive multiplexing, i.e. $N\ll 1$ generation schemes running in parallel.
However, it is also possible to use feedforward and multiplexing to further increase the success probability of this scheme, while reducing the resource requirement compared to having multiple parallel generation schemes.
Rather than building only one $k$-qubit ring, instead, we can use a ``factory'' that produces many $k$-qubit rings in parallel (thus multiplexing steps 1 and 2). Similarly, a second factory produces the ($n_1,n_2$) quantum parity codes (step 3-5). A third factory receives the output of these two factories and realizes in parallel steps 6 and 7, such that we finally produce $C_{(k, n_1, n_2)}$  redundantly-encoded graphs with a much higher success probability. In Appendix~\ref{sec_multiplexing}, we provide a detailed analysis and show that we can produce  $C_{(k, n_1, n_2)}$  redundantly-encoded graphs with arbitrarily high success probability $\tilde P_{C_{(k,n_1,n_2)}} \geq 1 - \varepsilon$. In addition, we show that we only need a linear multiplexing overhead to exponentially reduce the failure probability $\varepsilon$.

\section{Conclusion}

To conclude, we have proposed a hybrid generation scheme to produce arbitrary graph states. It is based on two complementary resources: the deterministic generation of entangled photons by single quantum emitters and linear-optic fusion gates. Contrary to other arbitrary graph state generation protocols, our work softens the requirement on the quantum emitters, which do not need to interact with each other nor with photons after their emission. We only need a controllable quantum emitter able to emit spin-entangled photons, a requirement already met with the present-day technology. A single non-interacting quantum emitter is not sufficient to produce deterministically arbitrary graph states, but we showed that it can produce a class of entangled photonic states that we call  redundantly-encoded linear clusters and which includes both linear cluster states and GHZ states.
To create arbitrary graph states using these non-interacting quantum emitters, we need to combine them with linear optic fusion gates. Even though linear-optic fusion gates are inherently probabilistic, we showed that we can still use them to produce graph states near-deterministically, in the case of negligible photon losses. Indeed, we introduced boosted fusion gates and showed that we can combine them well with  redundantly-encoded linear clusters to produce arbitrary graph states.
Finally, we illustrated this result with the generation of two important categories of graph states: multi-dimensional cluster states which are the universal resource for measurement-based quantum computing, and encoded ring graph states which are the resource state for fusion-based quantum computing (FBQC).

Moreover, we emphasize that our scheme is only using technologies that have already been experimentally demonstrated and should therefore be readily implementable with state-of-the-art quantum emitters~\cite{Cogan2021, Thomas2022, Coste2023} and linear optic circuits. Even though our hybrid generation proposal is only near-deterministic in a loss-less scenario, it should still provide a better generation rate than all-photonic (i.e., linear optics with postselection) schemes. Therefore, our work provides an immediate route towards intermediate-size graphs of complex topologies, thus potentially enabling  NISQ applications with photonic qubits.
Moreover, we propose an efficient strategy to produce the resource states necessary for fusion-based quantum computing, which shows that it could also have applications beyond NISQ, towards fault-tolerant quantum computing. Building on this work, an interesting direction for future research is to include the concept of recycling graph states~\cite{Pant2017}: after a failed fusion gate or boosted fusion gate, we can still reuse parts of the photonic states to further boost the generation rate and reduce the resource overhead. Combining this idea with hybrid graph state generation should facilitate the implementation of measurement-based photonic quantum information processing.

Even though the coherence time of state-of-the-art quantum emitters is typically short compared to some non-optically-active platforms, we should emphasize that they are only used to produce the photonic state and thus need to remain coherent only during the graph state generation steps. We expect that already with the current state-of-the-art~\cite{Cogan2021, Coste2023, Thomas2022}, we can, for example, produce 6-qubit ring graph states which still maintain quantum properties, given that we can in principle already produce linear cluster with longer entanglement length. With the continuing experimental development of better quantum emitter and integrated photonic platforms, we should reach higher and higher generation rates and better fidelities.
An interesting and important future direction for this work would be to devise an efficient error model which takes into account all the realistic imperfections occurring during this graph state generation (e.g.\@ spin decoherence, photon distinguishability, multi-photon emission, etc.) to quantify how they affect the properties of the photonic state and whether they can be handled fault-tolerantly.

\acknowledgments
This research was supported by the EU Horizon 2020 programme (GA 862035 QLUSTER).
S.E.E. also acknowledges the Virginia Commonwealth Cyber Initiative (CCI), an investment in the advancement of cyber R\&D, innovation, and workforce development (\href{www.cyberinitiative.org}{www.cyberinitiative.org}).

\appendix

\section{Comparison with other fusion schemes}\label{sec_comp}

In the following, we will compare quantitatively the performances of our scheme with ancilla-assisted fusion schemes~\cite{Grice2011, Ewert2014} and to the repeat-until-success scheme~\cite{Lim2005}.

\subsection{Ancilla-assisted scheme}
We focus on the protocols of Ref.~\cite{Grice2011} and Ref.~\cite{Ewert2014}. The former provides the best ancilla-assisted Bell state measurement performances in terms of resource usage. The ancilla photons that are used are prepared in entangled states which is why they exhibit better performances compared to schemes using disentangled ancilla photons such as in Ref.~\cite{Ewert2014}. Since our quantum emitters can also emit entangled photons, we consider that we have access to these entangled states ``for free'', and thus that we can prepare the ancilla photons of the Grice scheme efficiently~\cite{Grice2011}.
By using $k_G=2^{N+1} - 2$ ancilla photons, the success probability of the BSM is increased to $P_{s, G} = (k + 1) / (k + 2)$. However, these photons need all to be detected to herald a successful BSM. Therefore, the success probability of the Grice scheme is
$$ P_G(\eta, \eta_a, k_G) = P_{s,G} \eta^2 {\eta_a}^{k},$$
where $\eta$ is the collection efficiency of the photonic qubits, and $\eta_a$ is the collection efficiency of the ancilla photons.

We also consider the performances of the Ewert \& Van Loock scheme~\cite{Ewert2014} which are calculated similarly except that we have now $k_{\rm EvL}=2^{N+2} - 4$ and $P_{s, {\rm EvL}} = (k_{\rm EvL}+2) / (k_{\rm EvL}+4)$:
$$ P_{\rm EvL}(\eta, \eta_a, k_{\rm EvL}) = P_{s, {\rm EvL}} \eta^2 {\eta_a}^{k_{\rm EvL}}.$$

In the following, we will assume that $\eta_a = \eta$, which is a fair assumption if we are considering that the single-photon source and detector efficiencies are dominant over the channel losses, or that the ancilla photons experience the similar amount of channel losses than the photonic qubits.

\begin{figure}[ht]
  \centering
  \includegraphics[width=\columnwidth]{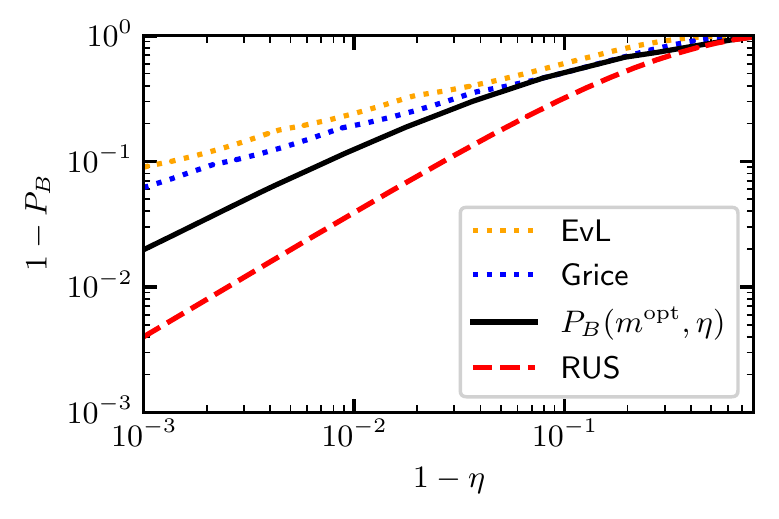}
  \caption{Performance comparison between our boosted fusion scheme ($P_{B}(\eta,m^{\rm opt})$), the repeat-until-success scheme (RUS) and ancilla-assisted fusion gates using Grice's scheme~\cite{Grice2011}.}
  \label{fig_boosted_fusion_comp}
\end{figure}

In that case, to make a fair comparison we indicate the number of photons involved in the ancilla-assisted fusion is $2m$ with $m = (k+1)/2$. In Fig.~\ref{fig_boosted_fusion_comp}, we compare the success probability of the optimal boosted fusion gate probability  $P_B(m^{\rm opt}, \eta)$ to the ancilla-assisted schemes $P_{G}(\eta, \eta, 2m-1)$ and $P_{\rm EvL}(\eta, \eta, 2m-1)$ (which we optimize to reach the best success probabilities). We observe that the boosted fusion gate performs systematically better that the ancilla-assisted schemes.

\subsection{Repeat-until-success scheme}

In the repeat-until-success (RUS) scheme, the fusion gates are executed until one succeeds (with probability $P_s = \eta^2/2$) or until at least one photon has been lost (failure probability $1 - \eta^2$). The fusion has failed in the latter case. The result of a given two-photon fusion can be indecisive with probability $P_i = \eta^2/2$ so that repeating until one has succeeded gives a success rate of 
$$P_{\rm RUS}(\eta) = \sum_{j = 1}^\infty P_s {P_i}^{j-1} = \sum_{j=1}^\infty \left(\frac{\eta^2}{2}\right)^{j} = \frac{\eta^2}{2 - \eta^2}.$$

As shown in Fig.~\ref{fig_boosted_fusion_comp}, the RUS scheme provides a better success rate than our boosted scheme but comes with the limitations discussed in the main text and is not applicable to our case.

\section{3D / $d$-dimensional cluster state.}
\label{sec_3D_cluster}
We now provide an analysis for a multi-dimensional cluster state generation which is a generalization of the 2D cluster state case.
This analysis works for arbitrary $d$-dimensional cluster state even though it seems technically challenging to go beyond 3D cluster states since implementing a $d$-dimensional cluster state requires an optical setup with at least $d-1$ connectivity.
In a $d$ dimensional cluster state, we can generate the first dimension deterministically using quantum emitters. In the 2D cluster state case, for the fusion gates connecting the second dimension, it was necessary to produce $2m$ additional photons at each vertex of this  redundantly-encoded linear cluster state. Similarly, each additional dimension requires here $2m$ additional photons to be generated at each vertex. Therefore, for a $d$-dimensional cluster state, the initial  redundantly-encoded cluster state should be a $n_1$-vertex  redundantly-encoded linear cluster state, where each vertex includes $1 + 2(d-1)m$ photons. If we want to produce a cluster of shape $n_1 \times n_2 \times ... \times n_d$ (described by the vector $\vec n_d=(n_1,...,n_d)$), we need to generate
\begin{equation}
  N_G(\vec n_d) = \sum_{i = 2}^d g(\vec n_d, i),
\end{equation}
fusion gates where
\begin{equation}
  g(\vec n_d, i) = (n_i - 1)\prod_{\substack{j=1 \\ j \neq i}}^d n_j
\end{equation}
is the number of fusion gates required in the $i^{\rm th}$ dimensions.
And the total success probability for the generation of such a cluster state is
\begin{equation}
  P_{\vec n_d}(\eta, m) = P_B(m,   \eta)^{N_G(\vec n_d)}.
\end{equation}

\section{Encoded ring graph states generation using large multiplexing.}
\label{sec_multiplexing}

In the following, we propose a more efficient strategy to create $k$-qubit ring graph states encoded with a ($n_1, n_2$) quantum parity code.
The general idea is to separate the generation scheme between three different ``factories'', denoted A, B, and C. Each factory is dedicated to a single operation: factory A is realizing the unencoded $k$-qubit rings, factory B is realizing the ($n_1, n_2$) codes and factory C takes as an input the $k$-qubit rings and the codes produced by factories A and B and create the encoded $k$-qubit ring graphs. The operation realized by each factory is multiplexed so that it can output its results with higher probability.

In factory A, $N_A$ quantum emitters (or a single sequential quantum emitter) are producing in parallel $k$-qubit ring graph states, with a success probability $p_a = P_B = P_B(m^{\rm (opt)}, \eta)$. At least $a$ $k$-qubit rings are created with a probability:
\begin{equation}
  P_A(\geq a) = P^{\geq a}(N_A, p_a) = 1 - \sum_{i=0}^{a-1} P^{(=i)}(N_A, p_a)
\end{equation}
where $P^{(=m)}(N, P)$ is the binomial probability
\begin{equation}
  P^{(=m)}(N, P) = {{N}\choose{m}} P^m (1-P)^{N-m}.
\end{equation}

In factory B, there are $N_B$ groups of $n_1 - 1$ quantum emitters and each group produce one ($n_1, n_2$) code.
Factory B is producing a ($n_1, n_2$) code with a success probability of $p_b = P_B^{(n_1 - 2)}$. Note that factory C requires sets of $k$ of such codes to be useful. Therefore, factory B produces $b$ sets of $k$ codes with probability:
\begin{equation}
  P_B(\geq b) = P^{\geq kb}(N_B, p_b).
\end{equation}

Factory C receive the graphs output from both factory A and B. If it receives $a$ $k$-qubit ring graph state from factory A and $b$ sets of $k$ codes from factory B, it can try to produce at most $c = \min(a, b)$ encoded graph states. The probability of receiving sufficient resources for at least $c$ encoded ring graph states is thus:
\begin{equation}
  P_{C,r}(\geq c) = P_A(\geq c) \times P_B(\geq c).
\end{equation}
Factory C thus receive exactly sufficient resources to try to produce $c$ encoded ring graph states with probability $P_{C,r}(=c) = P_{C,r}(\geq c) - P_{C,r}(\geq c+1)$.

 To produce an encoded ring graph state, each trial succeeds with a probability $p_c = (P_B \eta^2)^k$ and thus at least one encoded ring graph state is produced out of $c$ trial with probability
\begin{equation}
  P_{C,s}(=c) = 1 - (1 - p_c)^c.
\end{equation}

Finally, we can calculate the overall success probability of our three factories:
\begin{equation}
  P_S = \sum_{c = 1}^{\min(N_A, N_B)} P_{C,r}(=c) P_{C,s}(=c).
\end{equation}

Even in the presence of photon losses, this success probability of producing an encoded ring graph state tends to unity for increasing values of $N_A$ and $N_B$, .
Yet, this requires increasing the number of quantum emitters. Instead, we can optimize the overall success probability per number of quantum emitter used.

Now, we are looking for a good approximation of the amount of multiplexing needed to produce an encoding ring graph state with a success probability $P_S \approx 1 - \varepsilon$ with $0 < \varepsilon < 1$. Note that to obtain this success probability,  factory C typically needs $\hat c$ trials with:
\begin{equation}
  \hat c = \ceil*{\frac{\log(\varepsilon)}{\log(1 - p_c)}}
\end{equation}

We will thus get such a success probability if factories A and B produced reliably  $\hat c$ outputs. On average, they are producing on average $N_A p_a$ and $N_B p_b$ outputs respectively, so we should take $\hat N_A \approx \ceil*{\hat c / p_a}$ and $\hat N_B \approx \ceil*{k\hat c / p_b}$.  Therefore, we can reach arbitrarily-high success probability using this multiplexing strategy. Besides, we show that the failure probability $\varepsilon$ is reduced exponentially with the amount of multiplexing $N_A$, $N_B$.

\begin{figure}[!ht]
  \centering
  \includegraphics[width=\columnwidth]{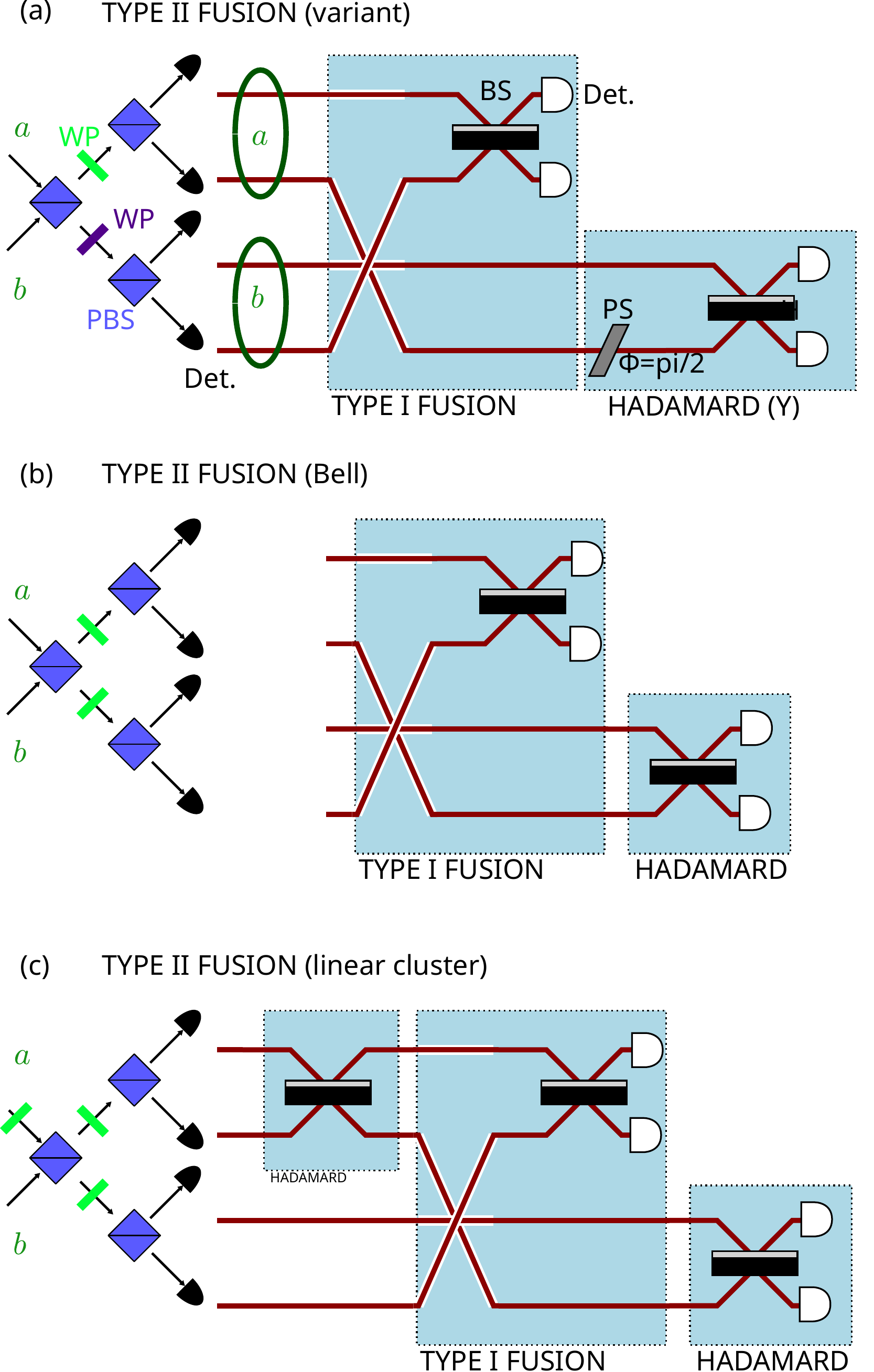}
  \caption{For each fusion gates, we provide a linear-optical setup operating either on polarization-encoded photonic qubits (left) or on path-encoded qubits (right), using Perceval~\cite{Heurtel2022}. (a) Type II fusion gate that we use in our boosted fusion gate scheme.  (b) Usual type II fusion gate that performs a Bell measurement in the $(\ket{00} \pm \ket{11})/\sqrt{2}$ basis. (c)  Variant of a type II fusion gate performing a measurement in the ``two-qubit linear cluster state'' basis $(\ket{0+} \pm \ket{1-})/\sqrt{2}$. (P)BS: (polarizing) beamsplitter, WP: waveplate, PS: phase shifter, Det: detector.}
  \label{fig_fusions}
\end{figure}

\section{Details on the type II fusion gates}\label{app_fusion}

In this section we provide details about the variant of the type II fusion gates that we are considering for boosted fusions that are described in in Fig.~\ref{fig_boosted_fusion}.
As shown in Fig.~\ref{fig_fusions}(a), its effect can be considered as a two-step process. The first step corresponds to a usual type I fusion gate described with the fusion operators $G_{I,i}(a, b) = \ket{0_a}\bra{0_a 0_b} + (-1)^i \ket{1_a} \bra{1_a 1_b}$, where $i=0,1$ corresponds to the measurement outcome, i.e. which detector has detected the photon. Note that $G_{I, 1}(a,b)= Z_a G_{I, 0}(a,b)$. The effect of a type I fusion gate is to fuse the two qubits $a$ and $b$ into 1.
Assuming we have two graphs $\ket{G_a}$, $\ket{G_b}$, containing respectively $a$ or $b$, upon successful type I fusion gate $G_{I, i}(a,b)$, we obtain the graph state $\ket{G_{a,b}}$ corresponding to a fusion of vertices $a$ and $b$:
$$G_{I, i}(a,b) \ket{G_a} \ket{G_b} = {Z_a}^i \ket{G_{a,b}}.$$
If two photons are detected on either the first 2 modes or the last two modes, a type I fusion gate heralds a failure, corresponding to two individual $Z$ measurements, thus disconnecting the qubits from the graph. The resulting graph corresponds thus to $\ket{G_{a} \backslash a} \ket{G_b \backslash b}$ up to local $Z$ rotation on the neighbors of $a$ and $b$.

The second step corresponds to a $Y_a$ single-qubit measurement, $P_{Y_a, j} = (I_a + (-1) ^j Y_a)/2$ where $j = 0, 1$ is the measurement outcome. The effect of a $Y_a$ measurement on a graph state corresponds to a local complementation $\tau_a(G)$ on the graph $G$ (up to local Clifford corrections $U_{a, j}$) followed by removing qubit $a$ from the graph~\cite{Hein2004, Hein2006}:
$$P_{Y_a, j} \ket{G} = U_{a, j} \ket{\tau_a(G) \backslash{a}},$$
with $U_{a, j} = \prod_{(a, v) \in E} \exp{\left[- i (-1)^j \pi Z_v /4\right]}$, a $(-1)^j\pi/2$ rotation along the $Z$-axis on all the neighbors of $a$.

Therefore, a successful type II fusion gate gives

\begin{align}
  G_{II, i,j} \ket{G_a} \ket{G_b} & = P_{Y_a, j} G_{I, i}(a,b) \ket{G_a} \ket{G_b}, \\
  &= P_{Y_a, j} {Z_a}^i \ket{G_{a,b}}, \\
  &= {Z_a}^i P_{Y_a, i \oplus j} \ket{G_{a,b}},\\
  & = {Z_a}^i U_{a, i \oplus j} \ket{\tau_a(G_{a,b}) \backslash{a}},
\end{align}
where we have used $P_{Y_a, j} {Z_a}^i = {Z_a}^i (I_a + (-1)^{i+j})/2 = {Z_a}^i P_{Y_a, i \oplus j}$.
The state is therefore local equivalent (up to $U_{a, i \oplus j}$) to the desired graph $\ket{\tau_a(G_{a,b}) \backslash{a}}$.

The interest of this variant of the type II fusion gates lies in using it for boosted fusion gates. In case of success, it creates the entanglement structure as shown in Fig.~\ref{fig_boosted_fusion} as expected. In case of detected failure, i.e.\@ the two photons are detected but do not herald a successful fusion, it corresponds to two individual $Z$ measurements, that simply disconnect the qubits from the graph. In comparison, if we use a type II fusion gate corresponding to a Bell state measurement (see Fig.~\ref{fig_fusions}(b)), i.e.\@ measuring $Z_a Z_b$ and $X_a X_b$, the failure case would be identical, but the success case would fuse the initial vertices of $a$ and $b$ and would thus give a different result than in Fig.~\ref{fig_boosted_fusion}. On the contrary, using a type II fusion corresponding to  measuring $X_a Z_b$ and $Z_a X_b$ (see Fig.~\ref{fig_fusions}(b)) would create an edge between the initial vertices of $a$ and $b$ in case of immediate success, but its detected failure would correspond to individual measurements $X_a$ and $Z_b$ (or conversely) which would modify the entanglement structure of the graph.

\bibliographystyle{unsrtnat}
\bibliography{bib_2}

\begin{thebibliography}{82}
\providecommand{\natexlab}[1]{#1}
\providecommand{\url}[1]{\texttt{#1}}
\expandafter\ifx\csname urlstyle\endcsname\relax
  \providecommand{\doi}[1]{doi: #1}\else
  \providecommand{\doi}{doi: \begingroup \urlstyle{rm}\Url}\fi

\bibitem[Zhong et~al.(2020)Zhong, Wang, Deng, Chen, Peng, Luo, Qin, Wu, Ding,
  Hu, et~al.]{Zhong2020}
Han-Sen Zhong, Hui Wang, Yu-Hao Deng, Ming-Cheng Chen, Li-Chao Peng, Yi-Han
  Luo, Jian Qin, Dian Wu, Xing Ding, Yi~Hu, et~al.
\newblock Quantum computational advantage using photons.
\newblock \emph{Science}, 370\penalty0 (6523):\penalty0 1460--1463, 2020.
\newblock \doi{10.1126/science.abe8770}.

\bibitem[Zhong et~al.(2021)Zhong, Deng, Qin, Wang, Chen, Peng, Luo, Wu, Gong,
  Su, et~al.]{Zhong2021}
Han-Sen Zhong, Yu-Hao Deng, Jian Qin, Hui Wang, Ming-Cheng Chen, Li-Chao Peng,
  Yi-Han Luo, Dian Wu, Si-Qiu Gong, Hao Su, et~al.
\newblock Phase-programmable gaussian boson sampling using stimulated squeezed
  light.
\newblock \emph{Physical review letters}, 127\penalty0 (18):\penalty0 180502,
  2021.
\newblock \doi{10.1103/PhysRevLett.127.180502}.

\bibitem[Arute et~al.(2019)Arute, Arya, Babbush, Bacon, Bardin, Barends,
  Biswas, Boixo, Brandao, Buell, et~al.]{Arute2019}
Frank Arute, Kunal Arya, Ryan Babbush, Dave Bacon, Joseph~C Bardin, Rami
  Barends, Rupak Biswas, Sergio Boixo, Fernando~GSL Brandao, David~A Buell,
  et~al.
\newblock Quantum supremacy using a programmable superconducting processor.
\newblock \emph{Nature}, 574\penalty0 (7779):\penalty0 505--510, 2019.
\newblock \doi{10.1038/s41586-019-1666-5}.

\bibitem[Knill et~al.(2001)Knill, Laflamme, and Milburn]{Knill2001}
Emanuel Knill, Raymond Laflamme, and Gerald~J Milburn.
\newblock A scheme for efficient quantum computation with linear optics.
\newblock \emph{Nature}, 409\penalty0 (6816):\penalty0 46--52, 2001.
\newblock \doi{10.1038/35051009}.

\bibitem[Raussendorf and Briegel(2001)]{Raussendorf2001}
Robert Raussendorf and Hans~J Briegel.
\newblock A one-way quantum computer.
\newblock \emph{Physical Review Letters}, 86\penalty0 (22):\penalty0 5188,
  2001.
\newblock \doi{10.1103/PhysRevLett.86.5188}.

\bibitem[Raussendorf et~al.(2006)Raussendorf, Harrington, and
  Goyal]{Raussendorf2006}
Robert Raussendorf, Jim Harrington, and Kovid Goyal.
\newblock A fault-tolerant one-way quantum computer.
\newblock \emph{Annals of physics}, 321\penalty0 (9):\penalty0 2242--2270,
  2006.
\newblock \doi{10.1016/j.aop.2006.01.012}.

\bibitem[Azuma et~al.(2015)Azuma, Tamaki, and Lo]{Azuma2015}
Koji Azuma, Kiyoshi Tamaki, and Hoi-Kwong Lo.
\newblock All-photonic quantum repeaters.
\newblock \emph{Nature communications}, 6:\penalty0 6787, 2015.
\newblock \doi{10.1038/ncomms7787}.

\bibitem[Ewert et~al.(2016)Ewert, Bergmann, and van Loock]{Ewert2016}
Fabian Ewert, Marcel Bergmann, and Peter van Loock.
\newblock Ultrafast long-distance quantum communication with static linear
  optics.
\newblock \emph{Physical review letters}, 117\penalty0 (21):\penalty0 210501,
  2016.
\newblock \doi{10.1103/PhysRevLett.117.210501}.

\bibitem[Lee et~al.(2019{\natexlab{a}})Lee, Ralph, and Jeong]{Lee2019b}
Seung-Woo Lee, Timothy~C Ralph, and Hyunseok Jeong.
\newblock Fundamental building block for all-optical scalable quantum networks.
\newblock \emph{Physical Review A}, 100\penalty0 (5):\penalty0 052303,
  2019{\natexlab{a}}.
\newblock \doi{10.1103/PhysRevA.100.052303}.

\bibitem[Hilaire et~al.(2021{\natexlab{a}})Hilaire, Barnes, Economou, and
  Grosshans]{Hilaire2021b}
Paul Hilaire, Edwin Barnes, Sophia~E. Economou, and Fr\'ed\'eric Grosshans.
\newblock Error-correcting entanglement swapping using a practical logical
  photon encoding.
\newblock \emph{Phys. Rev. A}, 104:\penalty0 052623, Nov 2021{\natexlab{a}}.
\newblock \doi{10.1103/PhysRevA.104.052623}.
\newblock URL \url{https://link.aps.org/doi/10.1103/PhysRevA.104.052623}.

\bibitem[Hilaire et~al.(2021{\natexlab{b}})Hilaire, Barnes, and
  Economou]{Hilaire2021}
Paul Hilaire, Edwin Barnes, and Sophia~E Economou.
\newblock Resource requirements for efficient quantum communication using
  all-photonic graph states generated from a few matter qubits.
\newblock \emph{Quantum}, 5:\penalty0 397, 2021{\natexlab{b}}.
\newblock \doi{10.22331/q-2021-02-15-397}.

\bibitem[Buterakos et~al.(2017)Buterakos, Barnes, and Economou]{Buterakos2017}
Donovan Buterakos, Edwin Barnes, and Sophia~E Economou.
\newblock Deterministic generation of all-photonic quantum repeaters from
  solid-state emitters.
\newblock \emph{Physical Review X}, 7\penalty0 (4):\penalty0 041023, 2017.
\newblock \doi{10.1103/PhysRevX.7.041023}.

\bibitem[Chan(2018)]{Chan2018}
Ming~Lai Chan.
\newblock Optimized protocol to create repeater graph states for all-photonic
  quantum repeater.
\newblock \emph{arXiv preprint arXiv:1811.10214}, 2018.
\newblock \doi{10.48550/arXiv.1811.10214}.

\bibitem[Russo et~al.(2018)Russo, Barnes, and Economou]{Russo2018}
Antonio Russo, Edwin Barnes, and Sophia~E Economou.
\newblock Photonic graph state generation from quantum dots and color centers
  for quantum communications.
\newblock \emph{Physical Review B}, 98\penalty0 (8):\penalty0 085303, 2018.
\newblock \doi{10.1103/PhysRevB.98.085303}.

\bibitem[Zhan and Sun(2020)]{Zhan2020}
Yuan Zhan and Shuo Sun.
\newblock Deterministic generation of loss-tolerant photonic cluster states
  with a single quantum emitter.
\newblock \emph{Physical Review Letters}, 125\penalty0 (22):\penalty0 223601,
  2020.
\newblock \doi{10.1103/PhysRevLett.125.223601}.

\bibitem[Browne and Rudolph(2005)]{Browne2005}
Daniel~E Browne and Terry Rudolph.
\newblock Resource-efficient linear optical quantum computation.
\newblock \emph{Physical Review Letters}, 95\penalty0 (1):\penalty0 010501,
  2005.
\newblock \doi{10.1103/PhysRevLett.95.010501}.

\bibitem[Rudolph(2017)]{Rudolph2017}
Terry Rudolph.
\newblock Why i am optimistic about the silicon-photonic route to quantum
  computing.
\newblock \emph{APL Photonics}, 2\penalty0 (3):\penalty0 030901, 2017.
\newblock \doi{10.1063/1.4976737}.

\bibitem[Bartolucci et~al.(2023)Bartolucci, Birchall, Bombin, Cable, Dawson,
  Gimeno-Segovia, Johnston, Kieling, Nickerson, Pant, et~al.]{Bartolucci2021}
Sara Bartolucci, Patrick Birchall, Hector Bombin, Hugo Cable, Chris Dawson,
  Mercedes Gimeno-Segovia, Eric Johnston, Konrad Kieling, Naomi Nickerson,
  Mihir Pant, et~al.
\newblock Fusion-based quantum computation.
\newblock \emph{Nature Communications}, 14\penalty0 (1):\penalty0 912, 2023.
\newblock \doi{10.1038/s41467-023-36493-1}.

\bibitem[Varnava et~al.(2008)Varnava, Browne, and Rudolph]{Varnava2008}
Michael Varnava, Daniel~E. Browne, and Terry Rudolph.
\newblock How good must single photon sources and detectors be for efficient
  linear optical quantum computation?
\newblock \emph{Physical Review Letters}, 100:\penalty0 060502, Feb 2008.
\newblock \doi{10.1103/PhysRevLett.100.060502}.
\newblock URL \url{http://link.aps.org/doi/10.1103/PhysRevLett.100.060502}.

\bibitem[Greganti et~al.(2021)Greganti, Demarie, Ringbauer, Jones, Saggio,
  Calafell, Rozema, Erhard, Meth, Postler, et~al.]{Greganti2021}
C~Greganti, TF~Demarie, M~Ringbauer, JA~Jones, V~Saggio, I~Alonso Calafell,
  LA~Rozema, A~Erhard, M~Meth, L~Postler, et~al.
\newblock Cross-verification of independent quantum devices.
\newblock \emph{Physical Review X}, 11\penalty0 (3):\penalty0 031049, 2021.
\newblock \doi{10.1103/PhysRevX.11.031049}.

\bibitem[Peruzzo et~al.(2014)Peruzzo, McClean, Shadbolt, Yung, Zhou, Love,
  Aspuru-Guzik, and O’brien]{Peruzzo2014}
Alberto Peruzzo, Jarrod McClean, Peter Shadbolt, Man-Hong Yung, Xiao-Qi Zhou,
  Peter~J Love, Al{\'a}n Aspuru-Guzik, and Jeremy~L O’brien.
\newblock A variational eigenvalue solver on a photonic quantum processor.
\newblock \emph{Nature communications}, 5\penalty0 (1):\penalty0 1--7, 2014.
\newblock \doi{10.1038/ncomms5213}.

\bibitem[Ferguson et~al.(2021)Ferguson, Dellantonio, Al~Balushi, Jansen,
  D{\"u}r, and Muschik]{Ferguson2021}
Ryan~R Ferguson, Luca Dellantonio, Abdulrahim Al~Balushi, Karl Jansen, Wolfgang
  D{\"u}r, and Christine~A Muschik.
\newblock Measurement-based variational quantum eigensolver.
\newblock \emph{Physical review letters}, 126\penalty0 (22):\penalty0 220501,
  2021.
\newblock \doi{10.1103/PhysRevLett.126.220501}.

\bibitem[Sch{\"o}n et~al.(2005)Sch{\"o}n, Solano, Verstraete, Cirac, and
  Wolf]{Schon2005}
Christian Sch{\"o}n, Enrique Solano, Frank Verstraete, J~Ignacio Cirac, and
  Michael~M Wolf.
\newblock Sequential generation of entangled multiqubit states.
\newblock \emph{Physical review letters}, 95\penalty0 (11):\penalty0 110503,
  2005.
\newblock \doi{10.1103/PhysRevLett.95.110503}.

\bibitem[Lindner and Rudolph(2009)]{Lindner2009}
Netanel~H Lindner and Terry Rudolph.
\newblock Proposal for pulsed on-demand sources of photonic cluster state
  strings.
\newblock \emph{Physical Review Letters}, 103\penalty0 (11):\penalty0 113602,
  2009.
\newblock \doi{10.1103/PhysRevLett.103.113602}.

\bibitem[Economou et~al.(2010)Economou, Lindner, and Rudolph]{Economou2010}
Sophia~E Economou, Netanel Lindner, and Terry Rudolph.
\newblock Optically generated 2-dimensional photonic cluster state from coupled
  quantum dots.
\newblock \emph{Physical review letters}, 105\penalty0 (9):\penalty0 093601,
  2010.
\newblock \doi{10.1103/PhysRevLett.105.093601}.

\bibitem[Russo et~al.(2019)Russo, Barnes, and Economou]{Russo2019}
Antonio Russo, Edwin Barnes, and Sophia~E Economou.
\newblock Generation of arbitrary all-photonic graph states from quantum
  emitters.
\newblock \emph{New Journal of Physics}, 21\penalty0 (5):\penalty0 055002,
  2019.
\newblock \doi{10.1088/1367-2630/ab193d}.

\bibitem[Gimeno-Segovia et~al.(2019)Gimeno-Segovia, Rudolph, and
  Economou]{Gimeno2019}
Mercedes Gimeno-Segovia, Terry Rudolph, and Sophia~E Economou.
\newblock Deterministic generation of large-scale entangled photonic cluster
  state from interacting solid state emitters.
\newblock \emph{Physical review letters}, 123\penalty0 (7):\penalty0 070501,
  2019.
\newblock \doi{10.1103/PhysRevLett.123.070501}.

\bibitem[Michaels et~al.(2021)Michaels, Mart{\'\i}nez, Debroux, Parker,
  Stramma, Huber, Purser, Atat{\"u}re, and Gangloff]{Michaels2021}
Cathryn~P Michaels, Jes{\'u}s~Arjona Mart{\'\i}nez, Romain Debroux, Ryan~A
  Parker, Alexander~M Stramma, Luca~I Huber, Carola~M Purser, Mete Atat{\"u}re,
  and Dorian~A Gangloff.
\newblock Multidimensional cluster states using a single spin-photon interface
  coupled strongly to an intrinsic nuclear register.
\newblock \emph{Quantum}, 5:\penalty0 565, 2021.
\newblock \doi{10.22331/q-2021-10-19-565}.

\bibitem[Li et~al.(2022)Li, Economou, and Barnes]{Li2022}
Bikun Li, Sophia~E Economou, and Edwin Barnes.
\newblock Photonic resource state generation from a minimal number of quantum
  emitters.
\newblock \emph{npj Quantum Information}, 8\penalty0 (1):\penalty0 1--7, 2022.
\newblock \doi{10.1038/s41534-022-00522-6}.

\bibitem[Pichler et~al.(2017)Pichler, Choi, Zoller, and Lukin]{Pichler2017}
Hannes Pichler, Soonwon Choi, Peter Zoller, and Mikhail~D Lukin.
\newblock Universal photonic quantum computation via time-delayed feedback.
\newblock \emph{Proceedings of the National Academy of Sciences}, 114\penalty0
  (43):\penalty0 11362--11367, 2017.
\newblock \doi{10.1073/pnas.1711003114}.

\bibitem[Wan et~al.(2021)Wan, Choi, Kim, Shutty, and Hayden]{Wan2020}
Kianna Wan, Soonwon Choi, Isaac~H Kim, Noah Shutty, and Patrick Hayden.
\newblock Fault-tolerant qubit from a constant number of components.
\newblock \emph{PRX Quantum}, 2\penalty0 (4):\penalty0 040345, 2021.
\newblock \doi{10.1103/PRXQuantum.2.040345}.

\bibitem[Shi and Waks(2021)]{Shi2021}
Yu~Shi and Edo Waks.
\newblock Deterministic generation of multidimensional photonic cluster states
  using time-delay feedback.
\newblock \emph{Physical Review A}, 104\penalty0 (1):\penalty0 013703, 2021.
\newblock \doi{10.1103/PhysRevA.104.013703}.

\bibitem[Zhong et~al.(2018)Zhong, Li, Li, Peng, Su, Hu, He, Ding, Zhang, Li,
  et~al.]{Zhong2018}
Han-Sen Zhong, Yuan Li, Wei Li, Li-Chao Peng, Zu-En Su, Yi~Hu, Yu-Ming He, Xing
  Ding, Weijun Zhang, Hao Li, et~al.
\newblock 12-photon entanglement and scalable scattershot boson sampling with
  optimal entangled-photon pairs from parametric down-conversion.
\newblock \emph{Physical review letters}, 121\penalty0 (25):\penalty0 250505,
  2018.
\newblock \doi{10.1103/PhysRevLett.121.250505}.

\bibitem[Istrati et~al.(2020)Istrati, Pilnyak, Loredo, Ant{\'o}n, Somaschi,
  Hilaire, Ollivier, Esmann, Cohen, Vidro, et~al.]{Istrati2020}
D~Istrati, Y~Pilnyak, JC~Loredo, C~Ant{\'o}n, N~Somaschi, P~Hilaire,
  H~Ollivier, M~Esmann, L~Cohen, L~Vidro, et~al.
\newblock Sequential generation of linear cluster states from a single photon
  emitter.
\newblock \emph{Nature communications}, 11\penalty0 (1):\penalty0 1--8, 2020.
\newblock \doi{10.1038/s41467-020-19341-4}.

\bibitem[Zhang et~al.(2022)Zhang, Liu, Li, Fei, Yin, Li, Liu, Mao, Chen, and
  Pan]{Zhang2022}
Rui Zhang, Li-Zheng Liu, Zheng-Da Li, Yue-Yang Fei, Xu-Fei Yin, Li~Li, Nai-Le
  Liu, Yingqiu Mao, Yu-Ao Chen, and Jian-Wei Pan.
\newblock Loss-tolerant all-photonic quantum repeater with generalized shor
  code.
\newblock \emph{Optica}, 9\penalty0 (2):\penalty0 152--158, 2022.
\newblock \doi{10.1364/OPTICA.439170}.

\bibitem[Schwartz et~al.(2016)Schwartz, Cogan, Schmidgall, Don, Gantz, Kenneth,
  Lindner, and Gershoni]{Schwartz2016}
Ido Schwartz, Dan Cogan, Emma~R Schmidgall, Yaroslav Don, Liron Gantz, Oded
  Kenneth, Netanel~H Lindner, and David Gershoni.
\newblock Deterministic generation of a cluster state of entangled photons.
\newblock \emph{Science}, 354:\penalty0 434--437, 2016.
\newblock \doi{DOI: 10.1126/science.aah4758}.

\bibitem[Besse et~al.(2020)Besse, Reuer, Collodo, Wulff, Wernli, Copetudo,
  Malz, Magnard, Akin, Gabureac, et~al.]{Besse2020}
Jean-Claude Besse, Kevin Reuer, Michele~C Collodo, Arne Wulff, Lucien Wernli,
  Adrian Copetudo, Daniel Malz, Paul Magnard, Abdulkadir Akin, Mihai Gabureac,
  et~al.
\newblock Realizing a deterministic source of multipartite-entangled photonic
  qubits.
\newblock \emph{Nature communications}, 11\penalty0 (1):\penalty0 1--6, 2020.
\newblock \doi{10.1038/s41467-020-18635-x}.

\bibitem[Cogan et~al.(2023)Cogan, Su, Kenneth, and Gershoni]{Cogan2021}
Dan Cogan, Zu-En Su, Oded Kenneth, and David Gershoni.
\newblock Deterministic generation of indistinguishable photons in a cluster
  state.
\newblock \emph{Nature Photonics}, pages 1--6, 2023.
\newblock \doi{10.1038/s41566-022-01152-2}.

\bibitem[Thomas et~al.(2022)Thomas, Ruscio, Morin, and Rempe]{Thomas2022}
Philip Thomas, Leonardo Ruscio, Olivier Morin, and Gerhard Rempe.
\newblock Efficient generation of entangled multiphoton graph states from a
  single atom.
\newblock \emph{Nature}, 608\penalty0 (7924):\penalty0 677--681, 2022.
\newblock \doi{10.1038/s41566-022-01152-2}.

\bibitem[Senellart et~al.(2017)Senellart, Solomon, and White]{Senellart2017}
Pascale Senellart, Glenn Solomon, and Andrew White.
\newblock High-performance semiconductor quantum-dot single-photon sources.
\newblock \emph{Nature nanotechnology}, 12\penalty0 (11):\penalty0 1026, 2017.
\newblock \doi{10.1038/nnano.2017.218}.

\bibitem[Jackson et~al.(2021)Jackson, Gangloff, Bodey, Zaporski, Bachorz,
  Clarke, Hugues, Le~Gall, and Atat{\"u}re]{Jackson2021}
Daniel~M Jackson, Dorian~A Gangloff, Jonathan~H Bodey, Leon Zaporski, Clara
  Bachorz, Edmund Clarke, Maxime Hugues, Claire Le~Gall, and Mete Atat{\"u}re.
\newblock Quantum sensing of a coherent single spin excitation in a nuclear
  ensemble.
\newblock \emph{Nature Physics}, pages 1--6, 2021.
\newblock \doi{10.1038/s41567-020-01161-4}.

\bibitem[Reiserer et~al.(2016)Reiserer, Kalb, Blok, van Bemmelen, Taminiau,
  Hanson, Twitchen, and Markham]{Reiserer2016}
Andreas Reiserer, Norbert Kalb, Machiel~S Blok, Koen~JM van Bemmelen, Tim~H
  Taminiau, Ronald Hanson, Daniel~J Twitchen, and Matthew Markham.
\newblock Robust quantum-network memory using decoherence-protected subspaces
  of nuclear spins.
\newblock \emph{Physical Review X}, 6\penalty0 (2):\penalty0 021040, 2016.
\newblock \doi{10.1103/PhysRevX.6.021040}.

\bibitem[Gottesman(1997)]{Gottesman1997}
Daniel Gottesman.
\newblock Stabilizer codes and quantum error correction.
\newblock \emph{arXiv preprint quant-ph/9705052}, 1997.
\newblock \doi{10.48550/arXiv.quant-ph/9705052}.

\bibitem[Nielsen and Chuang(2010)]{Nielsen2002}
Michael~A. Nielsen and Isaac~L. Chuang.
\newblock \emph{Quantum Computation and Quantum Information: 10th Anniversary
  Edition}.
\newblock Cambridge University Press, 2010.
\newblock \doi{10.1017/CBO9780511976667}.

\bibitem[Lee et~al.(2019{\natexlab{b}})Lee, Villa, Bennett, Stevenson, Ellis,
  Farrer, Ritchie, and Shields]{Lee2019}
JP~Lee, B~Villa, AJ~Bennett, RM~Stevenson, DJP Ellis, I~Farrer, DA~Ritchie, and
  AJ~Shields.
\newblock A quantum dot as a source of time-bin entangled multi-photon states.
\newblock \emph{Quantum Science and Technology}, 4\penalty0 (2):\penalty0
  025011, 2019{\natexlab{b}}.
\newblock \doi{10.1088/2058-9565/ab0a9b}.

\bibitem[Tiurev et~al.(2022)Tiurev, Appel, Mirambell, Lauritzen, Tiranov,
  Lodahl, and S{\o}rensen]{Tiurev2020}
Konstantin Tiurev, Martin~Hayhurst Appel, Pol~Llopart Mirambell, Mikkel~Bloch
  Lauritzen, Alexey Tiranov, Peter Lodahl, and Anders~S{\o}ndberg S{\o}rensen.
\newblock High-fidelity multiphoton-entangled cluster state with solid-state
  quantum emitters in photonic nanostructures.
\newblock \emph{Physical Review A}, 105\penalty0 (3):\penalty0 L030601, 2022.
\newblock \doi{10.1103/PhysRevA.105.L030601}.

\bibitem[Tiurev et~al.(2021)Tiurev, Mirambell, Lauritzen, Appel, Tiranov,
  Lodahl, and S{\o}rensen]{Tiurev2021}
Konstantin Tiurev, Pol~Llopart Mirambell, Mikkel~Bloch Lauritzen,
  Martin~Hayhurst Appel, Alexey Tiranov, Peter Lodahl, and Anders~S{\o}ndberg
  S{\o}rensen.
\newblock Fidelity of time-bin-entangled multiphoton states from a quantum
  emitter.
\newblock \emph{Physical Review A}, 104\penalty0 (5):\penalty0 052604, 2021.
\newblock \doi{10.1103/PhysRevA.104.052604}.

\bibitem[Bartolucci et~al.(2021)Bartolucci, Birchall, Gimeno-Segovia, Johnston,
  Kieling, Pant, Rudolph, Smith, Sparrow, and
  Vidrighin]{Bartolucci2021creation}
Sara Bartolucci, Patrick~M Birchall, Mercedes Gimeno-Segovia, Eric Johnston,
  Konrad Kieling, Mihir Pant, Terry Rudolph, Jake Smith, Chris Sparrow, and
  Mihai~D Vidrighin.
\newblock Creation of entangled photonic states using linear optics.
\newblock \emph{arXiv preprint arXiv:2106.13825}, 2021.
\newblock \doi{10.48550/arXiv.2106.13825}.

\bibitem[Pan et~al.(2012)Pan, Chen, Lu, Weinfurter, Zeilinger, and
  {\.Z}ukowski]{Pan2012}
Jian-Wei Pan, Zeng-Bing Chen, Chao-Yang Lu, Harald Weinfurter, Anton Zeilinger,
  and Marek {\.Z}ukowski.
\newblock Multiphoton entanglement and interferometry.
\newblock \emph{Reviews of Modern Physics}, 84\penalty0 (2):\penalty0 777,
  2012.
\newblock \doi{10.1103/RevModPhys.84.777}.

\bibitem[Grice(2011)]{Grice2011}
Warren~P Grice.
\newblock Arbitrarily complete bell-state measurement using only linear optical
  elements.
\newblock \emph{Physical Review A}, 84\penalty0 (4):\penalty0 042331, 2011.
\newblock \doi{10.1103/PhysRevA.84.042331}.

\bibitem[Ewert and van Loock(2014)]{Ewert2014}
Fabian Ewert and Peter van Loock.
\newblock 3/4-efficient bell measurement with passive linear optics and
  unentangled ancillae.
\newblock \emph{Physical review letters}, 113\penalty0 (14):\penalty0 140403,
  2014.
\newblock \doi{10.1103/PhysRevLett.113.140403}.

\bibitem[Olivo and Grosshans(2018)]{Olivo2018}
Andrea Olivo and Fr{\'e}d{\'e}ric Grosshans.
\newblock Ancilla-assisted linear optical bell measurements and their
  optimality.
\newblock \emph{Physical Review A}, 98\penalty0 (4):\penalty0 042323, 2018.
\newblock \doi{10.1103/PhysRevA.98.042323}.

\bibitem[Lim et~al.(2005)Lim, Beige, and Kwek]{Lim2005}
Yuan~Liang Lim, Almut Beige, and Leong~Chuan Kwek.
\newblock Repeat-until-success linear optics distributed quantum computing.
\newblock \emph{Physical review letters}, 95\penalty0 (3):\penalty0 030505,
  2005.
\newblock \doi{10.1103/PhysRevLett.95.030505}.

\bibitem[Barrett and Kok(2005)]{Barrett2005}
Sean~D Barrett and Pieter Kok.
\newblock Efficient high-fidelity quantum computation using matter qubits and
  linear optics.
\newblock \emph{Physical Review A}, 71\penalty0 (6):\penalty0 060310, 2005.
\newblock \doi{10.1103/PhysRevA.71.060310}.

\bibitem[Lim et~al.(2006)Lim, Barrett, Beige, Kok, and Kwek]{Lim2006}
Yuan~Liang Lim, Sean~D Barrett, Almut Beige, Pieter Kok, and Leong~Chuan Kwek.
\newblock Repeat-until-success quantum computing using stationary and flying
  qubits.
\newblock \emph{Physical Review A}, 73\penalty0 (1):\penalty0 012304, 2006.
\newblock \doi{10.1103/PhysRevA.73.012304}.

\bibitem[Pant et~al.(2017)Pant, Krovi, Englund, and Guha]{Pant2017}
Mihir Pant, Hari Krovi, Dirk Englund, and Saikat Guha.
\newblock Rate-distance tradeoff and resource costs for all-optical quantum
  repeaters.
\newblock \emph{Physical Review A}, 95\penalty0 (1):\penalty0 012304, 2017.
\newblock \doi{10.1103/PhysRevA.95.012304}.

\bibitem[Varnava et~al.(2006)Varnava, Browne, and Rudolph]{Varnava2006}
Michael Varnava, Daniel~E Browne, and Terry Rudolph.
\newblock Loss tolerance in one-way quantum computation via counterfactual
  error correction.
\newblock \emph{Physical review letters}, 97\penalty0 (12):\penalty0 120501,
  2006.
\newblock \doi{10.1103/PhysRevLett.97.120501}.

\bibitem[Bell et~al.(2022)Bell, Pettersson, and Paesani]{Bell2022}
Tom~J Bell, Love~A Pettersson, and Stefano Paesani.
\newblock Optimising graph codes for measurement-based loss tolerance.
\newblock \emph{arXiv preprint arXiv:2212.04834}, 2022.
\newblock \doi{10.48550/arXiv.2212.04834}.

\bibitem[Kambs and Becher(2018)]{Kambs2018}
Benjamin Kambs and Christoph Becher.
\newblock Limitations on the indistinguishability of photons from remote solid
  state sources.
\newblock \emph{New Journal of Physics}, 20\penalty0 (11):\penalty0 115003,
  2018.
\newblock \doi{10.1088/1367-2630/aaea99}.

\bibitem[Beugnon et~al.(2006)Beugnon, Jones, Dingjan, Darqui{\'e}, Messin,
  Browaeys, and Grangier]{Beugnon2006}
Jones Beugnon, Matthew~PA Jones, Jos Dingjan, Beno{\^\i}t Darqui{\'e},
  Ga{\"e}tan Messin, Antoine Browaeys, and Philippe Grangier.
\newblock Quantum interference between two single photons emitted by
  independently trapped atoms.
\newblock \emph{Nature}, 440\penalty0 (7085):\penalty0 779--782, 2006.
\newblock \doi{10.1038/nature04628}.

\bibitem[Maunz et~al.(2007)Maunz, Moehring, Olmschenk, Younge, Matsukevich, and
  Monroe]{Maunz2007}
Peter Maunz, DL~Moehring, S~Olmschenk, KC~Younge, DN~Matsukevich, and C~Monroe.
\newblock Quantum interference of photon pairs from two remote trapped atomic
  ions.
\newblock \emph{Nature Physics}, 3\penalty0 (8):\penalty0 538--541, 2007.
\newblock \doi{10.1038/nphys644}.

\bibitem[Patel et~al.(2010)Patel, Bennett, Farrer, Nicoll, Ritchie, and
  Shields]{Patel2010}
Raj~B Patel, Anthony~J Bennett, Ian Farrer, Christine~A Nicoll, David~A
  Ritchie, and Andrew~J Shields.
\newblock Two-photon interference of the emission from electrically tunable
  remote quantum dots.
\newblock \emph{Nature photonics}, 4\penalty0 (9):\penalty0 632--635, 2010.
\newblock \doi{10.1038/nphoton.2010.161}.

\bibitem[Giesz et~al.(2015)Giesz, Portalupi, Grange, Ant{\'o}n, De~Santis,
  Demory, Somaschi, Sagnes, Lema{\^\i}tre, Lanco, et~al.]{Giesz2015}
V~Giesz, SL~Portalupi, T~Grange, C~Ant{\'o}n, L~De~Santis, J~Demory,
  N~Somaschi, I~Sagnes, A~Lema{\^\i}tre, L~Lanco, et~al.
\newblock Cavity-enhanced two-photon interference using remote quantum dot
  sources.
\newblock \emph{Physical Review B}, 92\penalty0 (16):\penalty0 161302, 2015.
\newblock \doi{10.1103/PhysRevB.92.161302}.

\bibitem[Gold et~al.(2014)Gold, Thoma, Maier, Reitzenstein, Schneider,
  H{\"o}fling, and Kamp]{Gold2014}
P~Gold, A~Thoma, S~Maier, S~Reitzenstein, C~Schneider, S~H{\"o}fling, and
  M~Kamp.
\newblock Two-photon interference from remote quantum dots with inhomogeneously
  broadened linewidths.
\newblock \emph{Physical Review B}, 89\penalty0 (3):\penalty0 035313, 2014.
\newblock \doi{10.1103/PhysRevB.89.035313}.

\bibitem[Bernien et~al.(2012)Bernien, Childress, Robledo, Markham, Twitchen,
  and Hanson]{Bernien2012}
Hannes Bernien, Lilian Childress, Lucio Robledo, Matthew Markham, Daniel
  Twitchen, and Ronald Hanson.
\newblock Two-photon quantum interference from separate nitrogen vacancy
  centers in diamond.
\newblock \emph{Physical Review Letters}, 108\penalty0 (4):\penalty0 043604,
  2012.
\newblock \doi{10.1103/PhysRevLett.108.043604}.

\bibitem[Bernien et~al.(2013)Bernien, Hensen, Pfaff, Koolstra, Blok, Robledo,
  Taminiau, Markham, Twitchen, Childress, et~al.]{Bernien2013}
Hannes Bernien, Bas Hensen, Wolfgang Pfaff, Gerwin Koolstra, Machiel~S Blok,
  Lucio Robledo, Tim~H Taminiau, Matthew Markham, Daniel~J Twitchen, Lilian
  Childress, et~al.
\newblock Heralded entanglement between solid-state qubits separated by three
  metres.
\newblock \emph{Nature}, 497\penalty0 (7447):\penalty0 86--90, 2013.
\newblock \doi{10.1038/nature12016}.

\bibitem[Sipahigil et~al.(2014)Sipahigil, Jahnke, Rogers, Teraji, Isoya,
  Zibrov, Jelezko, and Lukin]{Sipahigil2014}
Alp Sipahigil, Kay~D Jahnke, Lachlan~J Rogers, Tokuyuki Teraji, Junichi Isoya,
  Alexander~S Zibrov, Fedor Jelezko, and Mikhail~D Lukin.
\newblock Indistinguishable photons from separated silicon-vacancy centers in
  diamond.
\newblock \emph{Physical review letters}, 113\penalty0 (11):\penalty0 113602,
  2014.
\newblock \doi{10.1103/PhysRevLett.113.113602}.

\bibitem[Stockill et~al.(2017)Stockill, Stanley, Huthmacher, Clarke, Hugues,
  Miller, Matthiesen, Le~Gall, and Atat{\"u}re]{Stockill2017}
Robert Stockill, MJ~Stanley, Lukas Huthmacher, E~Clarke, M~Hugues, AJ~Miller,
  C~Matthiesen, Claire Le~Gall, and Mete Atat{\"u}re.
\newblock Phase-tuned entangled state generation between distant spin qubits.
\newblock \emph{Physical review letters}, 119\penalty0 (1):\penalty0 010503,
  2017.
\newblock \doi{10.1103/PhysRevLett.119.010503}.

\bibitem[Delteil et~al.(2016)Delteil, Sun, Gao, Togan, Faelt, and
  Imamo{\u{g}}lu]{Delteil2016}
Aymeric Delteil, Zhe Sun, Wei-bo Gao, Emre Togan, Stefan Faelt, and Ata{\c{c}}
  Imamo{\u{g}}lu.
\newblock Generation of heralded entanglement between distant hole spins.
\newblock \emph{Nature Physics}, 12\penalty0 (3):\penalty0 218--223, 2016.
\newblock \doi{10.1038/nphys3605}.

\bibitem[Somaschi et~al.(2016)Somaschi, Giesz, De~Santis, Loredo, Almeida,
  Hornecker, Portalupi, Grange, Anton, Demory, et~al.]{Somaschi2016}
Niccolo Somaschi, Valerian Giesz, Lorenzo De~Santis, JC~Loredo, Marcelo~P
  Almeida, Gaston Hornecker, S~Luca Portalupi, Thomas Grange, Carlos Anton,
  Justin Demory, et~al.
\newblock Near-optimal single-photon sources in the solid state.
\newblock \emph{Nature Photonics}, 10\penalty0 (5):\penalty0 340--345, 2016.
\newblock \doi{10.1038/nphoton.2016.23}.

\bibitem[Ding et~al.(2016)Ding, He, Duan, Gregersen, Chen, Unsleber, Maier,
  Schneider, Kamp, H{\"o}fling, et~al.]{Ding2016}
Xing Ding, Yu~He, Z-C Duan, Niels Gregersen, M-C Chen, S~Unsleber, Sebastian
  Maier, Christian Schneider, Martin Kamp, Sven H{\"o}fling, et~al.
\newblock On-demand single photons with high extraction efficiency and
  near-unity indistinguishability from a resonantly driven quantum dot in a
  micropillar.
\newblock \emph{Physical review letters}, 116\penalty0 (2):\penalty0 020401,
  2016.
\newblock \doi{10.1103/PhysRevLett.116.020401}.

\bibitem[Uppu et~al.(2020)Uppu, Pedersen, Wang, Olesen, Papon, Zhou, Midolo,
  Scholz, Wieck, Ludwig, et~al.]{Uppu2020}
Ravitej Uppu, Freja~T Pedersen, Ying Wang, Cecilie~T Olesen, Camille Papon,
  Xiaoyan Zhou, Leonardo Midolo, Sven Scholz, Andreas~D Wieck, Arne Ludwig,
  et~al.
\newblock Scalable integrated single-photon source.
\newblock \emph{Science advances}, 6\penalty0 (50):\penalty0 eabc8268, 2020.
\newblock \doi{10.1126/sciadv.abc8268}.

\bibitem[Tomm et~al.(2021)Tomm, Javadi, Antoniadis, Najer, L{\"o}bl, Korsch,
  Schott, Valentin, Wieck, Ludwig, et~al.]{Tomm2021}
Natasha Tomm, Alisa Javadi, Nadia~Olympia Antoniadis, Daniel Najer,
  Matthias~Christian L{\"o}bl, Alexander~Rolf Korsch, R{\"u}diger Schott,
  Sascha~Ren{\'e} Valentin, Andreas~Dirk Wieck, Arne Ludwig, et~al.
\newblock A bright and fast source of coherent single photons.
\newblock \emph{Nature Nanotechnology}, pages 1--5, 2021.
\newblock \doi{10.1038/s41565-020-00831-x}.

\bibitem[Coste et~al.(2023)Coste, Fioretto, Belabas, Wein, Hilaire,
  Frantzeskakis, Gundin, Goes, Somaschi, Morassi, et~al.]{Coste2023}
N~Coste, DA~Fioretto, N~Belabas, SC~Wein, P~Hilaire, R~Frantzeskakis, M~Gundin,
  B~Goes, N~Somaschi, M~Morassi, et~al.
\newblock High-rate entanglement between a semiconductor spin and
  indistinguishable photons.
\newblock \emph{Nature Photonics}, pages 1--6, 2023.
\newblock \doi{10.1038/s41566-023-01186-0}.

\bibitem[Riedel et~al.(2017)Riedel, S{\"o}llner, Shields, Starosielec, Appel,
  Neu, Maletinsky, and Warburton]{Riedel2017}
Daniel Riedel, Immo S{\"o}llner, Brendan~J Shields, Sebastian Starosielec,
  Patrick Appel, Elke Neu, Patrick Maletinsky, and Richard~J Warburton.
\newblock Deterministic enhancement of coherent photon generation from a
  nitrogen-vacancy center in ultrapure diamond.
\newblock \emph{Physical Review X}, 7\penalty0 (3):\penalty0 031040, 2017.
\newblock \doi{10.1103/PhysRevX.7.031040}.

\bibitem[Zhang et~al.(2018)Zhang, Sun, Burek, Dory, Tzeng, Fischer, Kelaita,
  Lagoudakis, Radulaski, Shen, et~al.]{Zhang2018}
Jingyuan~Linda Zhang, Shuo Sun, Michael~J Burek, Constantin Dory, Yan-Kai
  Tzeng, Kevin~A Fischer, Yousif Kelaita, Konstantinos~G Lagoudakis, Marina
  Radulaski, Zhi-Xun Shen, et~al.
\newblock Strongly cavity-enhanced spontaneous emission from silicon-vacancy
  centers in diamond.
\newblock \emph{Nano letters}, 18\penalty0 (2):\penalty0 1360--1365, 2018.
\newblock \doi{10.1021/acs.nanolett.7b05075}.

\bibitem[Knall et~al.(2022)Knall, Knaut, Bekenstein, Assumpcao, Stroganov,
  Gong, Huan, Stas, Machielse, Chalupnik, et~al.]{Knall2022}
Erik~N Knall, Can~M Knaut, Rivka Bekenstein, Daniel~R Assumpcao, Pavel~L
  Stroganov, Wenjie Gong, Yan~Qi Huan, P-J Stas, Bartholomeus Machielse,
  Michelle Chalupnik, et~al.
\newblock Efficient source of shaped single photons based on an integrated
  diamond nanophotonic system.
\newblock \emph{Physical Review Letters}, 129\penalty0 (5):\penalty0 053603,
  2022.
\newblock \doi{10.1103/PhysRevLett.129.053603}.

\bibitem[Liu et~al.(2018)Liu, Brash, O’Hara, Martins, Phillips, Coles,
  Royall, Clarke, Bentham, Prtljaga, et~al.]{Liu2018}
Feng Liu, Alistair~J Brash, John O’Hara, Luis~MPP Martins, Catherine~L
  Phillips, Rikki~J Coles, Benjamin Royall, Edmund Clarke, Christopher Bentham,
  Nikola Prtljaga, et~al.
\newblock High purcell factor generation of indistinguishable on-chip single
  photons.
\newblock \emph{Nature nanotechnology}, 13\penalty0 (9):\penalty0 835--840,
  2018.
\newblock \doi{10.1038/s41565-018-0188-x}.

\bibitem[Ralph et~al.(2005)Ralph, Hayes, and Gilchrist]{Ralph2005}
Timothy~C Ralph, AJF Hayes, and Alexei Gilchrist.
\newblock Loss-tolerant optical qubits.
\newblock \emph{Physical review letters}, 95\penalty0 (10):\penalty0 100501,
  2005.
\newblock \doi{10.1103/PhysRevLett.95.100501}.

\bibitem[Heurtel et~al.(2023)Heurtel, Fyrillas, Gliniasty, Le~Bihan, Malherbe,
  Pailhas, Bertasi, Bourdoncle, Emeriau, Mezher, Music, Belabas, Valiron,
  Senellart, Mansfield, and Senellart]{Heurtel2022}
Nicolas Heurtel, Andreas Fyrillas, Gr{\'{e}}goire~de Gliniasty, Rapha{\"{e}}l
  Le~Bihan, S{\'{e}}bastien Malherbe, Marceau Pailhas, Eric Bertasi, Boris
  Bourdoncle, Pierre-Emmanuel Emeriau, Rawad Mezher, Luka Music, Nadia Belabas,
  Benoît Valiron, Pascale Senellart, Shane Mansfield, and Jean Senellart.
\newblock Perceval: {A} {S}oftware {P}latform for {D}iscrete {V}ariable
  {P}hotonic {Q}uantum {C}omputing.
\newblock \emph{{Quantum}}, 7:\penalty0 931, February 2023.
\newblock ISSN 2521-327X.
\newblock \doi{10.22331/q-2023-02-21-931}.
\newblock URL \url{https://doi.org/10.22331/q-2023-02-21-931}.

\bibitem[Hein et~al.(2004)Hein, Eisert, and Briegel]{Hein2004}
Marc Hein, Jens Eisert, and Hans~J Briegel.
\newblock Multiparty entanglement in graph states.
\newblock \emph{Physical Review A}, 69\penalty0 (6):\penalty0 062311, 2004.
\newblock \doi{10.1103/PhysRevA.69.062311}.

\bibitem[Hein et~al.(2006)Hein, D{\"u}r, Eisert, Raussendorf, Nest, and
  Briegel]{Hein2006}
Marc Hein, Wolfgang D{\"u}r, Jens Eisert, Robert Raussendorf, M~Nest, and H-J
  Briegel.
\newblock Entanglement in graph states and its applications.
\newblock \emph{arXiv preprint quant-ph/0602096}, 2006.
\newblock \doi{10.48550/arXiv.quant-ph/0602096}.

\end{thebibliography}

\end{document}